\def\Qem{{$Q_{\rm em}$}}

 \def\N{{$\cal N$}}
 \def\Z{{\bf Z}}
 \def\MG{{$M_{\rm GUT}$}}
\def\hf{\frac12}

\def\EE{E$_8\times$E$_8^\prime$}
\def\Uan{U(1)$_{\rm an}$}
\def\UglA{U(1)$_{\rm glA}$}
\def\UPQ{U(1)$_{\rm PQ}$}

\def\cagg{$c_{a\gamma\gamma}$}

\documentclass{JHEP3}
\usepackage{epsf,epsfig,subfigure,axodraw,graphicx,amsmath,amssymb}

\preprint{SNUTP 06-012}
\title{String compactification,
QCD axion and axion-photon-photon coupling
 }
\author{Kang-Sin Choi \\
Physikalisches Institut,
Universit\"at Bonn, Nussallee 12, D53115, Bonn, Germany
\\ E-mail: \email{kschoi@th.physik.uni-bonn.de}}
\author{Ian-Woo Kim \\
Department of Physics and Astronomy and Center for Theoretical
 Physics, Seoul National University, Seoul 151-747, Korea
\\ E-mail: \email{iwkim@phya.snu.ac.kr} }
\author{Jihn E.  Kim \\
Department of Physics and Astronomy and Center for Theoretical
 Physics, Seoul National University, Seoul 151-747, Korea
\\E-mail: \email{jekim@phyp.snu.ac.kr} }

\abstract{ It is pointed out that there exist a few problems to be
overcome toward an observable sub-eV QCD axion in superstring
compactification. We give a general expression for the axion decay
constant. For a large domain wall number $N_{DW}$, the axion decay
constant can be substantially lowered from a generic value of a
scalar singlet VEV. The Yukawa coupling structure in the recent
$\Z_{12-I}$ model is studied completely, including the needed
nonrenormalizable terms toward realistic quark and lepton masses. In
this model we find an approximate global symmetry and vacuum so that
a QCD axion results but its decay constant  is at the GUT scale. The
axion-photon-photon coupling is calculated for a realistic vacuum
satisfying the quark and lepton mass matrix conditions. It is the
first time calculation of \cagg\ in realistic string
compactifications: \cagg$=\frac{5}{3}-1.93\simeq -0.26$. }
\keywords{Yukawa couplings, Matter axion, String compactification}

\begin{document}

 \maketitle

\def\lsl{ l \hspace{-0.45 em}/}
\def\ksl{ k \hspace{-0.45 em}/}
\def\qsl{ q \hspace{-0.45 em}/}
\def\psl{ p \hspace{-0.45 em}/}
\def\ppsl{ p' \hspace{-0.70 em}/}
\def\dsl{ \partial \hspace{-0.5 em}/}
\def\Dsl{ D \hspace{-0.55 em}/}
\def\N{$\cal N$}
\def\tphi{\tilde\phi}

\section{Introduction and Summary on Superstring Axions}

The strong CP problem is, $\lq\lq$Why is the QCD vacuum angle
$\bar\theta_{\rm QCD}$ so small at $|\bar\theta_{\rm
QCD}|<10^{-9}$?" Otherwise, strong QCD interactions violate the CP
symmetry and then neutron will develop an electric dipole moment of
order $10^{-3}|\bar\theta_{\rm QCD}|\times$(charge radius of
neutron), and the present upper bound on the neutron electric dipole
moment $d_n< 0.63 \times 10^{-25} $e cm \cite{Harris:1999jx}
restricts $|\bar\theta_{\rm QCD}|<10^{-9}$. There are a few
solutions of the strong CP problem \cite{Kim:1986ax}: (i) Set
$\bar\theta_{\rm QCD}=0$ at tree level and guarantee that loop
effects are sufficiently small, (ii) $m_u=0$ method, and (iii) the
Peccei-Quinn (PQ) mechanism. Axion solutions which we discuss in
this paper belong to Case (iii). The PQ mechanism \cite{Peccei:1977hh}
introduces an anomalous (in the QCD gluon fields) global U(1)$_{\rm
PQ}$ symmetry. This must be an axial symmetry. Since quarks are
massive, the global U(1)$_{\rm PQ}$ symmetry must be spontaneously
broken, generating a Goldstone boson called axion \cite{PQWW}.
Currently, the phenomenologically allowable QCD axion is a very
light axion \cite{veryaxion}. A probable initial condition of the
axion decay constant allows the window, $10^{10}\ {\rm GeV}\le
F_a\le 10^{12}$ GeV. But, with the anthropic principle applied, the
upper bound can be further open \cite{WilAnth}.

Axion models in field theory are artificial in the sense that
the U(1)$_{\rm PQ}$ symmetry is given for the sake of the PQ
mechanism only. It is desirable if a consistent theory
with an ultraviolet completion gives a natural candidate for axion.
In this regards, we consider string models. If a string theory
predicts a phenomenologically allowable axion, this may be a key
prediction of string theory. This is welcome in view of the scarcity
of direct verifiable methods of $\lq$string' nature. In fact, one
attractive feature of string theory is the natural appearance of
axions from the antisymmetric tensor field $B_{MN}$. These
superstring axions are split into two categories: one is the
compactification scheme independent one called model-independent
axion (MI axion) $B_{\mu\nu}$ \cite{Witten:1984dg} and the other is the
compactification dependent one called model-dependent axion (MD axion) $B_{ij}$
\cite{Witten:1985fp}. However, these superstring axions are known to
have some problems. Derived from string theory, decay constants of
these axions are expected to be near the string scale $\sim 10^{16}$
GeV \cite{Choi:1985je}, outside the aforementioned axion window. For
MD axions, it is further known that the shift symmetries of MD
axions are broken at high energy scales \cite{Wen:1985jz} so that
they cannot be used as axions for rolling the vacuum angles to zero
\cite{Kim:1986ax}. However, one should not forget that such
superpotential of MD axions is a model-dependent statement
\cite{Polch}. Recently, there has been attempts to lower the decay
constants in string models or in higher dimensional models
\cite{Conlon:2006tq,Svrcek:2006yi,Kim:2006aq,Flacke:2006ad,Choi:2006za}.
One approach is a large volume compactification to lower the
fundamental scale \cite{Conlon:2006tq}. Another approach is using the
warped geometry to lower the scale. In heterotic flux
compactification, MD axions can be localized at a vanishing cycle
which is warped due to the flux so that we may have a small axion
decay constant \cite{Kim:2006aq}. Similar setup has been discussed
in a higher-dimensional model\cite{Flacke:2006ad}. In other
contexts, an axion in the Kachru-Kallosh-Linde-Trivedi setup has also
been discussed in \cite{Choi:2006za}.

In this paper, we restrict the discussion on superstring axions to
heterotic string only, but the generic problem of axion mixing is
present in any superstring axion models. For the MI axion, the decay
constant is near the scale $\sim 10^{16}$ GeV \cite{Choi:1985je}. Some
string compactifications such as the simple compactifications of
Refs. \cite{Candelas:1985en,Dixon:1985jw} do not lead to an
anomalous U(1), in which
case the MI axion is harmful \cite{Choi:1985je}. Later, it was found that
some string compactifications lead to an anomalous U(1) gauge
symmetry \cite{Barr:1985hk}. This anomalous symmetry can be gauged
owing to the Green-Schwarz mechanism by which the antisymmetric
tensor field $B_{\mu\nu}$ transforms nonlinearly under the U(1)
symmetry so as to cancel the anomaly \cite{Barr:1985hk}. This anomalous
U(1) gauge symmetry is a subgroup of SO(32) or of \EE. The
Green-Schwarz term \cite{Green:1984sg} makes it possible for  this anomalous
U(1) gauge boson to absorb the MI axion and become massive. Below
this gauge boson mass scale $\sim 10^{16}$ GeV, there results a
global symmetry \Uan. This is a kind of 't Hooft mechanism
\cite{'tHooft:1971rn}. Under this circumstance, if some scalar field
carrying \Uan\ charges develop VEVs around $\sim 10^{11}$ GeV, then
we obtain a harmless very light axion  by the PQ mechanism
\cite{Peccei:1977hh}. However, for this \Uan, the story is not that simple.
Most  light fields carry nonvanishing \Uan\ charges, and if one is forced
to give GUT scale VEVs to those
singlets for successful Yukawa coupling textures, then the \Uan\
breaking scale must be the GUT scale and the resulting axion is
again harmful. The QCD axion with this \Uan\ has been extensively
studied without any hidden sector confining force in \cite{Kim:1988dd},
where the Yukawa coupling textures were not used. Moreover, this
model is phenomenologically unsatisfactory since
$\sin^2\theta_W\ll\frac38$.

Most string compactifications need another confining force in the
hidden sector for the purpose of introducing supersymmetry breaking.
Then, we need an additional global symmetry to settle both QCD and
hidden sector $\theta$s. As mentioned above, only the \Uan, related
to the MI axion, is a good one to consider below the string scale.
Except this \Uan, there is no global continuous symmetry resulting
from string theory. For an additional global symmetry, we are only
at a disposal of approximate global symmetries from string
compactification. It is better for this approximate global symmetry
to be broken by a sufficiently high dimensional operators in the
superpotential so that the symmetry breaking superpotential is
negligible compared to the axion potential derived from anomalies.
This idea was examined in a SUGRA field theory model with a discrete
$\Z_3\times\Z_3$ symmetry \cite{Lazarides:1985bj}. But it has not been
studied in string compactifications. A discrete symmetry is a good
guideline to make approximate symmetry violating terms appear at
higher orders. If PQ symmetry breaking scales are around the
intermediate scale, it has been known that the PQ symmetry breaking
superpotential must be forbidden up to $D=9$ terms
\cite{Kamionkowski:1992mf}.
But this statement is an oversimplified one because higher order
terms can involve some scalar fields developing small VEVs or even
not developing any VEV. So, the PQ symmetry violating terms in the
superpotential must be checked in model-by-model bases. Note that
for the terms breaking the global symmetry to appear at a
sufficiently higher order, we need a large $N$, presumably $N=12$,
in the $\Z_N$ orbifold compactifications.

By the way, the MSSMs from superstring need Yukawa
couplings beyond cubic terms \cite{Kim:2006hv,Kobayashi:2004ya}.
So far, there has not
appeared any model where only cubic terms are sufficient to give all
the needed Yukawa couplings. So it is not unreasonable to require
that for realistic quark and lepton masses superstring models need
nonrenormalizable terms beyond cubic terms.

Thus, to realize a QCD axion in string compactifications, we must
satisfy the following conditions:
\begin{itemize}
\item One must work in a phenomenologically successful string
derived model.
\item One needs an additional confining group beyond QCD, and an
approximate global symmetry must be introduced. The MD axions cannot
be used since the world sheet instanton effects violate the shift
symmetries of the MD axions.
\item The Yukawa coupling structure must be studied carefully to
derive the approximate global symmetry.
\item With two axions, the axion mixing effect must be clarified.
\end{itemize}
So far, there has not appeared any literature satisfying all the
above conditions. In this paper, we try to explore the possibility
of satisfying all these conditions in the recently proposed string
MSSM \cite{Kim:2006hv}. We find a vacuum satisfying all of these
conditions, but the QCD axion decay constant falls in the GUT scale
region.

We emphasize the importance of the last condition which has been
overlooked in many superstring axion models. It is studied in Refs.
\cite{Kim:1998kx,Kim:2006aq}. There is the {\it cross theorem on axion
potential heights and decay constants}. It is for the case of two
$\theta$s with a complete mixing of two axions by the higher
potential that {\it the smaller decay constant corresponds to the
higher height of the axion potentials and the larger decay constant
corresponds to the lower height of the axion potentials}. It is
shown for the case of two axions with the MI axion and the MD axion
corresponding to the breathing mode moduli \cite{Choi:1985bz} where
the couplings are $\sim a_1(F\tilde F+F^\prime\tilde
F^\prime)+a_2(F\tilde F-F^\prime\tilde F^\prime)$.
In fact, the axion mixing occurs when both axions couple to the
anomaly which give rise to the higher potential from instanton.
 In the example of \cite{Choi:1985bz}, two anomalies couple
to both axions and hence the condition for the theorem is satisfied.
In such cases,  if the hidden sector axion potential is higher than the QCD
axion potential (as one might guess), then the decay constant of the
QCD axion is the GUT scale.
If we realize that the decay constant of the QCD axion falls around
$\sim 10^{11}$ GeV, then it will be observable by axion detection
experiments like the CAST of CERN \cite{Zioutas:2004hi}.

After solving the strong CP problem, we can consider a related, the
so-called $\mu$ problem \cite{Kim:1983dt}, derivable from the global
symmetries of the MSSM. The common origin of a very light axion
scale and supergravity scale was pointed out early
\cite{Kim:1983ia}. Phenomenologically, the strength of the $\mu$
parameter in $-\mu H_uH_d$ is required to be at the electroweak
scale. The first $\mu$ problem is why it is forbidden at the Planck
scale. The second $\mu$ problem is why it is of order the
electroweak scale. There have been suggestions that if it is
forbidden at the GUT scale, it is expected to be generated at the
electroweak scale in supergravity models \cite{Giudice:1988yz} and
in string models \cite{Casas:1992mk}. However, these solutions need
some symmetry anyway from the Yukawa coupling structure to forbid it
at scales below the GUT scale \cite{KimNilles93}. If many singlet VEVs are
required at the GUT scale, then forbidding just $H_uH_d$ in the
renormalizable superpotential as done in \cite{Casas:1992mk} is not enough
to exclude the $\mu$ term at the GUT scale.

In this paper, in addition we try to calculate the
axion-photon-photon coupling \cagg\ in a realistic superstring
model. For the MI axion, a mechanism to find out the global \Uan\
was explicitly given before \cite{Kim:1988dd} but a calculation of \cagg\
in that model has not been meaningful because the model is not
realistic due to $\sin^2\theta_W\ll\frac38$ and there is no
additional confining force for supersymmetry breaking. Now we
obtained a realistic superstring standard model in a $\Z_{12-I}$
construction \cite{Kim:2006hv} and here we can calculate \cagg\ if
there exists a QCD axion.

We may introduce two scales (at \MG\ and $10^{11}$ GeV) for breaking
\Uan\ and an approximate global symmetry \UglA. The resulting
approximate global current must be anomalous in the gluon fields
and/or hidden confining gauge fields. The axion mixing is studied
with the hidden sector axion potential.

 In this paper, we pick up an approximate global
symmetry \UglA\ in addition to \Uan. We use a computer program
which is based on the Gauss elimination algorithm to study all
possible U(1)  symmetries from any order of Yukawa couplings. A
complete analysis has been done for all Yukawa couplings up to $D=9$
superpotential terms. Unfortunately, we could not realize the QCD
axion with $F_a\sim 10^{11}$ GeV but at the GUT scale. The axion
energy crisis may be resolved by the anthropic principle
\cite{WilAnth}. We calculate the phenomenologically important
axion-photon-photon coupling constant \cagg, i.e. $\tilde
c_{a\gamma\gamma}=\frac53$ or \cagg $\simeq -0.26$. This is the
first calculation of \cagg\ from a realistic string
compactification.

This paper is organized as follows. In Sec. \ref{sec:axion}, we
summarize briefly axion physics related to the calculation of the
axion decay constant. This involves a discussion on the domain wall
number of discrete vacua. In Sec. \ref{sec:z12model}, we succinctly
present the recent $\Z_{12-I}$ orbifold model \cite{Kim:2006hv}
toward a computer input for the matter fields. In Sec.
\ref{sec:u1s}, we summarize the Yukawa coupling structure and find
the anomalous U(1)$_{\rm an}$ symmetry from \EE\ and an approximate
anomalous global symmetry U(1)$_{\rm glA}$  so that two $\theta$s
can be settled to zero via the PQ mechanism. In Sec. \ref{sec:agg},
we calculate the axion-photon-photon coupling by calculating
anomalies. It is compared to the recent CAST experiment bound
\cite{Zioutas:2004hi}. Sec. \ref{sec:Conc} is a conclusion.

\section{Axion, Domain Walls and Axion Decay Constant}
\label{sec:axion}

The QCD axion $a$ is defined to be a pseudoscalar particle coupling
to the gluon anomaly,
\begin{equation}
\frac{a}{32\pi^2 F_a}G_{\mu\nu}\tilde G^{\mu\nu}\equiv
\frac{a}{F_a}\{ G\tilde G\}\label{axdef}
\end{equation}
where $\tilde G^{\mu\nu}$ is the dual of $G^{\mu\nu}$, the gluon
kinetic term is $(1/4g_c^2)G_{\mu\nu}G^{\mu\nu}$ and $F_a$ is the
axion decay constant. It is assumed that there exists the canonical
kinetic energy term of $a$, i.e. $\frac12\partial_\mu a\partial^\mu
a$, and there is no potential for the axion except that derivable
from Eq. (\ref{axdef}). Then, the minimum of the axion potential is
at $\langle a\rangle=0$ \cite{Peccei:1977hh,Vafa:1984xg} which solves
the strong CP
problem cosmologically in the present universe \cite{Preskill:1982cy}.
We do not repeat the explanation of the solution of strong CP problem
using the axion in this section. For a complete review, refer to
\cite{Kim:1986ax}. Here, we discuss some subtlety of determining
the axion decay constant due to the remaining discrete group after
U(1)$_{\rm PQ}$ breakdown.

Because of the quantization of the integral of $G\tilde G$, the
axion potential is periodic with the periodicity $2\pi F_a$. In this
form Eq. (\ref{axdef}), the fundamental region of the axion vacua is
$[0,2\pi F_a]$, because $\bar\theta=[0,2\pi]$. Namely, starting from
the vacuum $\langle a\rangle=0$, the next vacuum occurs by shifting
$\langle a\rangle\to \langle a\rangle+2\pi F_a$. But the vacuum at
$\langle a\rangle=2\pi F_a$ may not be the same vacuum as the
$\langle a\rangle=0$ vacuum, but returns to the $\langle a\rangle=0$
vacuum only after the shift $a\to a+2\pi N_{DW}F_a$. Then there are
degenerate vacua whose number is called the domain wall number
$N_{DW}$ \cite{Sikivie:1982qv}. The axion embedded in some fields may not
return to its original value when one shifts $a\to a+2\pi F_a$,
which is the reason for the appearance of degenerate vacua. There is
another degeneracy from the scalar field space. Suppose, the axion
$a$ embedded in the phase of a scalar field as
$\phi=[{(v+\rho)/\sqrt2}]e^{i a/v}$ where $\phi$ carries $N$ units
of PQ charge. The field $\phi$ returns to its original value by
shifting $a\to a+2\pi v$. But the phase factor becomes identity for
$N$ distinct points of $0\le a<2\pi v$. This is related to the
domain wall number calculation. If we define the PQ charge $Q$ such
that $\phi$ carries one unit of $Q$ and the anomaly calculation
leads to $\partial_\mu J^\mu_{\rm PQ}=n\{G\tilde G\}$, the phase
$a$ couples to the anomaly as $\frac{a}{v}n\{G\tilde G\}$. Since
$\phi$ is defined such that it comes to the original value by the PQ
transformation or the shift of $a$ by $a\to a+2\pi v$, viz. $Q=1$,
this interaction defines $F_a=v/n$. Here, the domain wall number is
$N_{DW}=n$.

\begin{figure}[t]
\begin{center}
\begin{picture}(400,115)(0,0)
\LongArrow(50,90)(50,105)\Text(50,112)[c]{phase}
\LongArrow(160,20)(175,20)\Text(180,20)[l]{anomaly}
 \Text(50,20)[c]{$\bullet$} \BCirc(50,50){2}
 \Text(75,50)[c]{$\bullet$} \Text(100,80)[c]{$\bullet$}
  \Text(100,20)[c]{$\bullet$}
 \Text(125,50)[c]{$\bullet$}
 \Text(150,80)[c]{$\bullet$}
 \Text(150,20)[c]{$\bullet$}
 {\SetWidth{1}
\Line(50,20)(100,80)  \Line(100,20)(150,80)
\DashLine(75,20)(125,80){4}
\DashLine(125,20)(150,50){4}\DashLine(50,50)(75,80){4} }
\Text(100,0)[c]{(a)}

\Line(250,20)(250,80)\Line(250,80)(350,80)
\Line(350,20)(350,80)\Line(250,20)(350,20)

 \Text(50,80)[c]{$\bullet$}
 \BCirc(100,50){2}
 \BCirc(150,50){2} 
\BCirc(75,20){2} 
\BCirc(125,20){2} 
\BCirc(75,80){2}
 \BCirc(125,80){2}

\Text(300,0)[c]{(b)} \Text(251,20)[c]{$\bullet$}
\Text(350,80)[c]{$\bullet$}
 {\SetWidth{1.2}\LongArrow(250,20)(300,50) }
\end{picture}
\caption{Illustration of degenerate vacua for $N_1=4, N_2=2$.
Bullets and circles correspond to minima of the axion potential. (a)
$N_{DW}$ is two, the vacua connected by solid and dashed lines. (b)
The fundamental length (the arrow) in the fundamental region
corresponds to the decay constant which is $F_a=V/N_{DW}\Gamma$.
}\label{DWfig}
\end{center}
\end{figure}
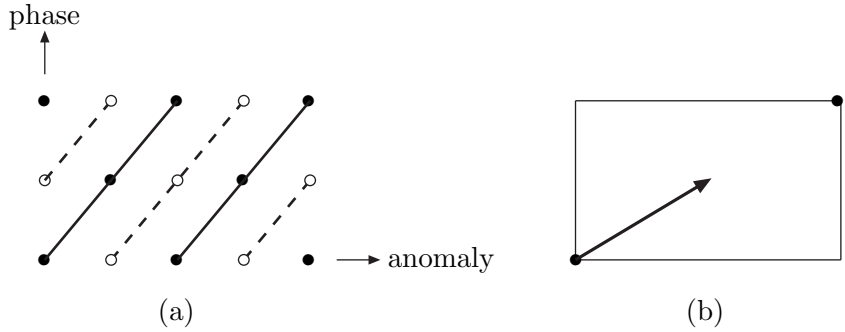

On the other hand, if we are required (by the PQ charges of the
other fields) to define the PQ charge of $\phi$ as $\Gamma$ which is
a positive integer greater than 1, then $\partial_\mu J^\mu_{\rm
PQ}=\Gamma n\{G\tilde G\}$, and we expect the axion decay constant
$F_a=v/n\Gamma$. But the domain wall number defined from vacuum
structure is a topological one and still we should obtain
$N_{DW}=n$. This situation is illustrated for $N_1=n\Gamma=4$ and
$N_2=\Gamma=2$, i.e. with $n=2$ in Fig. \ref{DWfig}.
Certainly, the minima of the axion potential has two distinct sets
in Fig. \ref{DWfig}(a), the set of bullets and the set of circles,
represented by the connections with solid and dashed lines,
respectively. These two sets agree with our original  domain wall
number, $N_{DW}=n=2$. This domain wall number is obtained from $N_1$
(the coefficient of anomaly) by dividing it with the greatest common
divisor of $N_1$ and $N_2$. In Fig. \ref{DWfig}(b), we show how our
$F_a$ is related to the original VEV.

Having specified the domain wall number, it is important to pick up
the properly normalized axion field toward calculating the axion
decay constant. Consider two fields $\phi_1$ and $\phi_2$. In the
unitary gauge, we can express the Goldstone bosons in the phases of
$\phi_1$ and $\phi_2$ as
\begin{align}
\phi_1=\frac{V_1+\rho_1}{\sqrt2}e^{iA_1/V_1},\quad
\phi_2=\frac{V_2+\rho_2}{\sqrt2}e^{iA_2/V_2}
\end{align}
whose PQ charges are $\Gamma_1$ and $\Gamma_2$, respectively. Now we
can restrict the phase fields $A_1=[0,\frac{1}{\Gamma_1} V_1]$ and
$A_2=[0,\frac{1}{\Gamma_2} V_2]$. If we are sitting in one domain,
say on the solid line (or on the dashed line) of Fig. \ref{DWfig}
for reinterpreting the figure with $A_1$ and $A_2$ directions with
$\Gamma_1=4$ and $\Gamma_2=2$, $A_1$ and $A_2$ completely define the
allowed field space in that domain. Thus, we can represent the
fields in one domain as
\begin{align}
\phi_1=\frac{V_1+\rho_1}{\sqrt2}e^{iA_1/(V_1/4)},
\quad A_1=[0,\frac{V_1}{2}]\\
\phi_2=\frac{V_2+\rho_1}{\sqrt2}e^{iA_2/(V_2/2)},\quad
A_2=[0,\frac{V_2}{2}].
\end{align}
The relevant VEV for axion is $\sqrt{4^2V_1^2+2^2V_2^2}$. The axion
direction in terms of $\lq A$'s is
$\theta_A=\frac{4A_1+2A_2}{\sqrt{4^2V_1^2+2^2V_2^2}}$. Suppose that
the gluon anomaly were just one unit of $\{G\tilde G\}$. Then, the
axion coupling is given by
\begin{align}
\frac{4A_1+2A_2}{\sqrt{4^2V_1^2+2^2V_2^2}}\{G\tilde G\}
\label{Axion1}
\end{align}
With the shifts of $A_1\to A_1+2\pi V_1$ and $A_2\to A_2+2\pi V_2$,
viz. Fig. \ref{DWfig}, the field space ends at the other domain,
i.e. $A_i$ shifted such that $\theta$ changed by $4\pi$. So to
maintain the fields in the same domain with $\theta=[0,2\pi)$, we
use $A_i= 2a_i$ such that $a_i\to a_i+2\pi F_i$. Thus, we have
$F_i=V_i/2$. Thus, the coupling (\ref{Axion1}) can be rewritten as
\begin{align}
\frac{2(4a_1+ 2a_2)}{\sqrt{4^2V_1^2+2^2V_2^2}}\{G\tilde G\}
 =2 ({\rm dimensionless\ axion\ component\ in\ the\ phase})
  \{G\tilde G\}\label{DWphase2}
\end{align}
where the overall factor 2 appears as the greatest common divisor
(GCD) of $\Gamma_1$ and $\Gamma_2$. For one field VEV dominating the
other, e.g. the DFSZ model with a large singlet VEV $\langle
s^0\rangle$ and a small doublet VEV, the dimensionless axion
component is in fact $a_1/\langle s^0\rangle$ and the axion decay
constant is $F_a=\langle s^0\rangle/2$.

If we consider the domain wall number from the gluon anomaly, that
should be considered also and an appropriate factor must be
multiplied to Eq. (\ref{DWphase2}). If the anomaly coefficient
$N(\rm anom.)$ from the gluon anomaly and $N({\rm phase})$ from a
multitude of phases are relatively prime, then the domain wall
number is simply $N({\rm anom.})$. But if they are not relatively
prime, we must divide  it by the greatest common divisor of
$N({\rm anom.})$ and $N({\rm phase})$. Therefore, a general
expression for the axion coupling can be written as
\begin{align}
\frac{N_{DW}\cdot D_{\rm GCD}({\rm phase})}{V_{\rm max}}\ a\{G\tilde
G\}\label{DWphaseN}
\end{align}
where $D_{\rm GCD}$ is the greatest common divisor of PQ charges
of the relevant VEV-carrying fields, and the topological number
$N_{DW}$ is calculated as in Fig. \ref{DWfig}(a) from the
coefficient of the anomaly and the phase degeneracy, i.e.
$N_{DW}=N({\rm anom.})/D_{\rm GCD}(N({\rm anom.}),N({\rm phase}))$.
The normalized axion field is defined as
\begin{equation}
\frac{a}{V_{\rm max}}\equiv \frac{\sum_i\Gamma_i
a_i}{\sqrt{\sum\Gamma_i^2 V_i^2}}\label{DefAxion}
\end{equation}
where $\Gamma_i$ is the U(1)$_{\rm PQ}$ charge of $s_i$, $V_i$ is
the VEV of scalar field $\langle s_i\rangle$, and $V_{\rm max}$ is a
typical scale of VEVs breaking the global symmetry. In many cases it
is the maximum value among $\langle s_i\rangle$. For example, for
the case $V_{\rm max}\gg V_i(i\ne {\rm max})$ and also for $V_{\rm
max}=V_i ({\rm for\ any\ }i)$. Then the axion decay constant is
\begin{equation}
F_a=\frac{ V_{\rm max}}{N_{DW}\cdot D_{\rm GCD}}\ .\label{AxDecay}
\end{equation}
Note that if $N_{DW}\cdot D_{\rm GCD}$ were large, then $F_a$ can be
lowered significantly compared to the PQ symmetry breaking scale.

The discussion of this section will be used in Sec. \ref{sec:agg} in
estimating the decay constant of the QCD axion from superstring.

\section{$\Z_{12-I}$ Orbifold Model}\label{sec:z12model}

In this section, we briefly summarize the key points of a realistic
string-derived minimal supersymmetric standard model (MSSM) from a
$\Z_{12-I}$ orbifold compactification \cite{Kim:2006hv}. Here, the low
energy effective theory contains the MSSM. We discuss vacuum
configuration of the MSSM singlets for realistic low energy
phenomenology, especially toward the absence of exotic matter
spectra and  general formulae for realistic Yukawa couplings. These
formulae are used in the program to generate all Yukawa couplings.
We pick up the anomalous gauged U(1), \Uan, and approximate global
U(1)s which appear in general in lower dimensional superpotential
terms. For gauged U(1)s, the gauge U(1) directions can be assigned
in the \EE\ group space but for the approximate global U(1)s it
cannot be represented in that way. The only method for approximate
global U(1)s is to list the U(1) charges of the matter fields.

\subsection{Flipped SU(5) matter spectrum and
Yukawa couplings}

In the orbifold compactification in heterotic string theories, the
action of twisting is represented as a shift vecdtor $V^I ( I = 1 ,
2, \dots , 16)$ and Wilson Lines $a_i^I ( i = 1 ,2, \dots, 6 ; I = 1
, 2, \dots, 16 )$. In the ${\bf Z}_{12-I}$ orbifold model discussed in
\cite{Kim:2006hv}, we choose SU(3)$\times$SO(9) lattice for toroidal
compactification and
\begin{align}
&V=\textstyle\left( \frac14~ \frac14~ \frac14~ \frac{1}{4}~ \frac14~
\frac{5}{12}~\frac{6}{12}~  0~ \right)\left( \frac{2}{12}~
\frac{2}{12}~ 0~ 0~0~0~0~0 \right)\label{shiftV}\\
&a_3=a_4=\textstyle\left( 0^5~0~ \frac{-1}{3}~ \frac{1}{3}~
\right)\left( 0~ 0~\frac{2}{3}~0^5 \right)\label{Wilson3}\\
&a_1=a_2=a_5=a_6=0.\nonumber
\end{align}
Then, below the compactification scale one obtains the following
unbroken gauge group
\begin{equation}
[SU(5)\times U(1)_X\times U(1)^3]\times [SU(2)\times SO(10)\times
 U(1)^2]^\prime.
\end{equation}
The spectra for the unbroken subgroup SU(5)$\times $U(1)$_X$ is in
fact the flipped SU(5) model \cite{Kim:2006hv}. The matter
representations under the nonabelian gauge groups SU(5),
SU(2)$^\prime$ and SO(10)$^\prime$ are simply determined by
embedding the momenta $P$ of \EE\ group space into the weight space
of the corresponding subgroup. The U(1) charges for matter fields
are determined by
\begin{eqnarray}
q_i = Z_i  \cdot ( P + k (V + m_f a_3 ) ),
\end{eqnarray}
where $i = X, 1 , 2, 3, 4 , 5$ denotes each U(1) gauge group and $k$
corresponds to the orbifold twist number of $k$-th twisted sector
and $m_f$ is the Wilson line twist number applicable only to the
second two-torus, viz. (\ref{Wilson3}). $Z_i$ are the following
E$_8\times$E$_8^\prime$ weights
\begin{align}
&Z_1=(2,2,2,2,2;0^3)(0^8)'
\\
&Z_2=(0^5;1,0,0)(0^8)'
\\
&Z_3=(0^5;0,1,0)(0^8)'
\\
&Z_4=(0^5;0,0,1)(0^8)'
\\
&Z_5=(0^8)(1,1;0;0^5)'
\\
&Z_6=(0^8)(0,0;1;0^5)'.
\end{align}
We will rearrange these U(1)s  when we consider the gauge anomalies
of the model completely.

The 4D Lorentz symmetry representation is solely determined by the right-mover
oscillators. Since we have supersymmetry in the low energy limit, we
consider either the fermion spectrum or the boson spectrum.
Consider the fermion spectrum. It is determined by the Ramond sector
vacuum of the right movers which is SO(8) spinor $s=(s_0,\tilde
s)=\{\pm\hf,\pm\hf, \pm\hf,\pm\hf\}$ with an even number of minus
signs. The 4D chirality $\chi$ is the first component of $s$, and we
call $\chi = \frac{1}{2}$ the right-handed field and $\chi =
-\frac{1}{2}$ the left-handed field.

Matter spectrum must satisfy the mass-shell conditions :
\begin{eqnarray}
&&  \frac{(P+kV)^2}{2} + \sum_j N_j^L \tilde{\phi}_j - \tilde{c} = 0,  \\
&&  \frac{(s+ k\phi_s)^2}{2}+ \sum_j N_j^R \tilde{\phi}_j - c = 0,
\end{eqnarray}
where $j$ runs over $\{ 1,2,3, \bar{1}, \bar{2}, \bar{3}\}$ and
$\tilde{\phi}_i \equiv k \phi_i$ and $ \tilde{\phi}_{\bar{i}}
\equiv -k \phi_i$
where $\equiv$ denotes one plus the maximum integer smaller than
the original
real number. Here $N_j^L$ and $N_j^R$ are the oscillator numbers,
and $c$ and
$\tilde{c}$ are the zero point energy of the right mover and left mover,
respectively.

By the modular invariance, the matter spectrum satisfying the above
mass shell conditions can be further projected out and have multiple
copies of themselves. This can be summarized by one-loop partition
function. The correct formulae are reviewed in \cite{ChoiKim06},
\begin{eqnarray}
{\mathcal P}_k = \frac{1}{N}\sum_{l = 0 }^{N}
\tilde{\chi} ( \theta^k , \theta^l )
\frac{1}{N_W} \sum_{f=0}^{N_W - 1} e^{2 \pi i l\Theta_f} ,
\end{eqnarray}
where $N$ is the order $N$ in the $\Z_N$ orbifold, $N_W$ is the
order of the Wilson line, 3 in our case, and
\begin{eqnarray}
\Theta_f = \sum_j (N^L_j - N^R_j ) \hat{\phi}_j
- \frac{k}{2} (V_f^2 - \phi_s^2 ) + (P+kV_f)\cdot V
- (\tilde{s} + k \phi_s )\cdot \phi_s,
\end{eqnarray}
where $\hat{\phi}_i = \phi_{s  i} {\rm sgn} (\tilde{\phi}_i)$ and
 $V_f = ( V+ m_f a )$.
Here, $\tilde{\chi}(\theta^k, \theta^l)$ is the degeneracy factor
summarized in \cite{Kim:2006hv}. ${\mathcal P}_k$ is interpreted as
the multiplicity of the spectrum. Thus, if ${\mathcal P}_k = 0$,
then the field is projected out by the modular invariance. Multiple
${\mathcal P}_k$ implies that the same matter spectra occur at
different orbifold fixed points. We summarize the resultant
 matter spectrum of this model in Tabel \ref{spectrum}
 and \ref{spectrum2}.

\TABLE{
\caption{ \label{spectrum} Spectrum of the Kim-Kyae $\Z_{12-I}$
model: 1. U sector and $T6, T1, T7$ sectors. }
\begin{tabular}{|c|cc|c||c|cc|c|}
\hline
\hline
 Name & $SU(5)_{U(1)_X}$ & $U(1)^5$ & P
&  Name & $SU(5)_{U(1)_X}$ & $U(1)^5$ & P\\
      &  $\times SU(2)'$ & &
&      &  $\times SU(2)'$ & &  \\
\hline
\multicolumn{4}{|c||}{ Untwisted } & \multicolumn{4}{|c|}{ Twisted 6} \\
\hline
 $U3_5$ & $( {\bf 5}_{\bf 3}, 1 )$ &$(6,6,6,0,0)$ & 1 &
        $T6_5$ & $({\bf 5}_{\bf 3}, 1)$ & (6,0,0,0,0) & 2 \\
        & & & &
       $T6_{\bar{5}}$ & $( \bar{{\bf 5}}_{\bf -3}, 1 )$ &(-6, 0, 0, 0, 0) & 2 \\
 $U3_{\overline{10}}$ & $(\overline{{\bf 10}}_{\bf -1} , 1)$
& $(6,-6,-6,0,0)$ &1 &
          $T6_{\overline{{10}}}$ &  $(\overline{{\bf 10}}_{\bf -1},1)$ &
 (6,0,0,0,0) & 4 \\
       & & & &
          $T6_{{10}}$ &   $({\bf 10}_{\bf 1}, 1)$ &
         (-6,0,0,0,0) & 3 \\
  $U3_1$ & $({\bf 1}_{\bf -5} , 1)$ & (6,6,6,0,0) & 1 &
          $T6_{1I}, T6_{1J}$ & $( {\bf 1}_{\bf \mp 5},1)$ &
          $(\pm 6,0,0,0,0)$ & 2,2 \\
  $U2_{\bar{5}}$ & $(\overline{{\bf 5}}_{\bf 2} , 1)$& (-12, 0, 0, 0, 0 ) & 1 &
          $h_1,\bar{h}_1$ & $({\bf 1}_{\bf 0},1)$& $(0,\pm 6,\pm 6, 0, 0)$ & 4,2 \\
  $s^u$ & $({\bf 1}_{\bf 0},1)$ &  (0,0,0,24,0) & 1 &
          $h_2, \bar{h}_2$ & $({\bf 1}_{\bf 0},1)$ &
          $(0,\pm 6, \pm 6, 0, 0 )$ & 2,3 \\
  $U1_{\bar{10}}$ & $(\overline{{\bf 10}}_{-1} , 1)$ & ( 6, 6, 6, 0 , 0 ) & 1 &
          $h_3, \bar{h}_3$ & $({\bf 1}_{\bf 0},1)$& $(0,\pm 6, \pm 6, 0, 0)$ & 2,4 \\
  $U1_5$ & $({\bf 5}_3 , 1)$ & (6,-6,-6,0,0 ) & 1 &
          $h_4, \bar{h}_4$ & $({\bf 1}_{\bf 0},1)$& $(0,\pm 6, \pm 6, 0, 0)$ & 3,2 \\
  $U1_1$ & $({\bf 1}_{\bf -5}, 1)$ & (6,-6, -6,0,0) & 1 &
          & & & \\
\hline
\multicolumn{4}{|c||}{ Twisted 1 } & \multicolumn{4}{|c|}{ Twisted 7} \\
\hline
 $T1_{\bar{5}A}$ & $(\overline{{\bf 5}}_{\bf -\frac{1}{2}},1)$ &
(-7,6,0,4,0) & 1  &
         $T7_{\bar{5}A}$ & $(\overline{{\bf 5}}_{\bf -\frac{1}{2}},1)$
         & (-7,0,-6,4,0) & 1\\
 $T1_{\bar{5}B}$ & $(\overline{{\bf 5}}_{\bf -\frac{1}{2}},1)$ &
(5,-6,0,4,0) & 1 &
         $T7_{\bar{5}B}$ & $(\overline{{\bf 5}}_{\bf -\frac{1}{2}},1)$
         & (5,0,6,4,0)  & 1\\
 $T1_5$ & $({\bf 5}_{\bf \frac{1}{2}},1)$ & (-1,0,6,4,0) &1 &
         $T7_5$ & $({\bf 5}_{\bf \frac{1}{2}} ,1 )$ & (-1,-6,0,4,0) & 1\\
 $T1_{1A}$& $({\bf 1}_{\bf-\frac{5}{2}},1)$ & (-7,-6,0,4,0) & 1 &
         $T7_{1A}$ & $({\bf 1}_{\bf -\frac{5}{2}},1)$ & (-7, 0, 6, 4, 0 ) & 1 \\
 $T1_{1B,C}$& $({\bf 1}_{\bf-\frac{5}{2}},1)$ & (5,6,0,4,0) &1 &
         $T7_{1B,C}$ & $({\bf 1}_{\bf -\frac{5}{2}},1)$ & (5,0,-6,4,0) & 1 \\
 $T1_{1D}$& $({\bf 1}_{\bf \frac{5}{2}},1)$ & (11,0,6,4,0) & 1 &
         $T7_{1D}$ & $({\bf 1}_{\bf \frac{5}{2}},1)$ & (11,-6,0,4,0) & 1 \\
 $T1_{1E,F,G}$& $({\bf 1}_{\bf \frac{5}{2}},1)$ & (-1,0,-6,4,0) & 1 &
         $T7_{1E,F,G}$ & $({\bf 1}_{\bf \frac{5}{2}},1)$ & (-1,6,0,4,0)&1\\
 $\delta_1$ & $({\bf 1}_{\bf -\frac{5}{2}},2)$ & (5,2,4,-8,-4) & 1&
         $\delta_3$ & $({\bf 1}_{\bf -\frac{5}{2}},2)$ & (5,-4,-2,-8,-4) & 1\\
 $T1^+_{1B}$ & $({\bf 1}_{\bf -\frac{5}{2}},1)$ & (5,2,4,4,8) &1 &
         $T7^+_{1B}$ & $({\bf 1}_{\bf -\frac{5}{2}},1)$ & (5,-4,-2,4,8) & 1 \\
 $\delta_2$ & $({\bf 1}_{\bf \frac{5}{2}},2)$ & (-1,-4,-2,-8,-4) &1 &
         $\delta_4$ & $({\bf 1}_{\bf \frac{5}{2}},2)$ & (-1,2,4,-8,-4) & 1 \\
 $T1^+_{1D}$ & $({\bf 1}_{\bf\frac{5}{2}},1)$ & (-1,-4,-2,4,8) & 1 &
          $T7^+_{1D}$ & $({\bf 1}_{\bf \frac{5}{2}},1)$ & (-1,2,4,4,8) & 1 \\
$T1^-_5$  & $({\bf 5}_{\bf \frac{1}{2}},1)$ & (-1,4,2,4,-8) & 1 &
         $T7^-_5$ & $({\bf 5}_{\frac{1}{2}},1)$ & (-1,-2,-4,4,-8 ) & 1 \\
 $T1^-_{1A}$ & $({\bf 1}_{\bf -\frac{5}{2}},1)$ & (5,-2,8,4,-8) & 1&
         $T7^-_{1A}$ & $({\bf 1}_{\bf -\frac{5}{2}},1)$ & (5,-8,2,4,-8) & 1 \\
$T1^-_{1B}$ & $({\bf 1}_{\bf -\frac{5}{2}},1)$ & (-7,-2,-4,4,-8) & 1 &
         $T7^-_{1B}$ & $({\bf 1}_{\bf -\frac{5}{2}},1)$ & (-7,4,2,4,-8) & 1 \\
$T1^-_{1C}$ & $({\bf 1}_{\bf \frac{5}{2}},1)$ & (-1,-8,2,4,-8) &1 &
         $T7^-_{1C}$ & $({\bf 1}_{\bf \frac{5}{2}},1)$ & (-1,-2,8,4,-8) & 1 \\
\hline
\end{tabular}
}

\TABLE{
\caption{ \label{spectrum2} Spectrum of the Kim-Kyae $Z_{12}$ model,
 {\it continued}:
2. $T2$ and $T4$ sectors.
Here, $O\bar{16}$ and $O10$ mean $\overline{\bf 16}$ representation and
${\bf 10}$ representation of SO(10)$'$, respectively.}
\begin{tabular}{|c|cc|c||c|cc|c|}
\hline
\hline
 Name & $SU(5)_{U(1)_X}$ & $U(1)^5$ & P
&  Name & $SU(5)_{U(1)_X}$ & $U(1)^5$ & P\\
      &  $\times SU(2)'$ & &
&      &  $\times SU(2)'$ & &  \\
\hline
\multicolumn{4}{|c||}{ Twisted 2 } &
\multicolumn{4}{|c|}{ Twisted 4 } \\
\hline
 $T2_5$ & $({\bf 5}_{\bf 3} ,1 )$ & $(-2,0,0,8,0)$ & 1 &
             $T4_5$ & $({\bf 5}_{\bf -2},1)$ & (-4,0,0,-8,0) & 3 \\
 $T2_1$ & $({\bf 1}_{\bf -5},1)$ & $(-2,0,0,8,0)$ & 1 &
             $T4_{\bar{5}}$ & $(\overline{{\bf 5}}_{\bf 2},1)$ & (-4,0,0,-8,0) & 2 \\
 $C^0_{1,2}$ & $({\bf 1}_{\bf 0},1)$ & (4,6,-6,8,0) & 1 &
             $s^0_{1,2,3}$ & $({\bf 1}_{\bf 0},1)$ & (8,0,0,-8,0) & 2 \\
 $C^0_{3,4}$ & $({\bf 1}_{\bf 0},1)$ & (4,-6,6,8,0) & 1 &
             $s^0_4$ & $({\bf 1}_{\bf 0},1)$ & (8,0,0,16,0) & 2 \\
 $C^0_5$ & $({\bf 1}_{\bf 0},1)$ & (-8,6,6,8,0) & 1 &
             $s^0_5$ & $({\bf 1}_{\bf 0},1)$ & (-4,12,0,-8,0) & 2 \\
 $C^0_6$ & $({\bf 1}_{\bf 0},1)$ & (-8,-6,-6,8,0) & 1 &
             $s^0_6$ & $({\bf 1}_{\bf 0},1)$ & (-4,-12,0,-8,0) & 2 \\
 $C^0_7$ & $({\bf 1}_{\bf 0},1)$ & (4,-6,6,-16,0) & 1 &
             $s^0_7$ & $({\bf 1}_{\bf 0},1)$ & (-4,0,12,-8,0 ) & 2 \\
 & & & &
             $s^0_8$ & $({\bf 1}_{\bf 0},1)$ & (-4,0,-12,-8,0 ) & 2 \\
 $T2^+_{O\overline{16}}$ &$({\bf 1}_{\bf 0},1)$
& (4,-2,2,-4,-2)  &  1 &
             $d^+_1$ & $({\bf 1}_{\bf 0},2)$ & (8,-4,4,4,-4) & 2 \\
 $D^+_{1,2}$ & $({\bf 1}_{\bf 0},2)$ & (4,2,-2,-4,-4) & 1&
             $d^+_2$ & $({\bf 1}_{\bf 0},2)$ & (-4,8,4,4,-4) & 2 \\
 $C^+_{1,2}$ & $({\bf 1}_{\bf 0},1)$ & (4,2,-2,8,8) & 1 &
             $d^+_3$ & $({\bf 1}_{\bf 0},2)$ & (-4,-4,-8,4,-4) & 2 \\
 & & & &
             $s^+_1$ & $({\bf 1}_{\bf 0},1)$ & (8,-4,4,-8,8) & 3 \\
 & & & &
             $s^+_2$ & $({\bf 1}_{\bf 0},1)$ & (-4,8,4,-8,8) & 2 \\
 & & & &
             $s^+_3$ & $({\bf 1}_{\bf 0},1)$ & (-4,-4,-8,-8,8) & 2 \\
 $T2^-_{O10}$&$({\bf 1}_{\bf 0},1)$
 & (4,2,-2,8,-4)& 1  &
             $d^-_1$ & $({\bf 1}_{\bf 0},2)$ & (8,4,-4,4,4) & 2 \\
 $D^-_{1,2}$ & $({\bf 1}_{\bf 0},2)$ & (4,-2,2,-4,4) & 1 &
             $d^-_2$ & $({\bf 1}_{\bf 0},2)$ & (-4,-8,-4,4,4) & 2 \\
 $C^-_{1,2}$ & $({\bf 1}_{\bf 0},1)$ & (4,-2,2,8,-8) & 1 &
             $d^-_3$ & $({\bf 1}_{\bf 0},2)$ & (-4,4,8,4,4) & 2 \\
 $C^-_3$       &$({\bf 1}_{\bf 0},1)$ &(4,2,-2,-16,8) &  1  &
             $s^-_1$ & $({\bf 1}_{\bf 0},1)$ & (8,4,-4,-8,-8)& 3 \\
            &           &              &   &
             $s^-_2$ & $({\bf 1}_{\bf 0},1)$ & (-4,-8,-4,-8,-8) & 2 \\
            &           &              &   &
             $s^-_3$ & $({\bf 1}_{\bf 0},1)$ & (-4,4,8,-8,-8) & 2 \\
 \hline
\end{tabular}
}

We summarize the left-handed  matter field
spectra,\footnote{We show only the
left-handed spectra. The right-handed spectra are just their PCT
conjugates.} by classifying them under the flipped-SU(5)  and
SO(10)$'$,
\begin{equation}
U\ :\ ({\bf 1}_{\bf -5}+{\bf {5}}_{\bf 3}+\overline{\bf 10}_{\bf
-1})_{U_3},\quad (\overline{\bf 5}_{\bf 2})_{U_2},\quad({\bf
1}_{\bf -5}+{\bf {5}}_{\bf 3}+\overline{\bf 10}_{\bf
-1})_{U_1},\quad ({\bf 1_0})_{U_2},\label{Cuntw}
\end{equation}
from the untwisted sector, and
\begin{align}
&T_6\ :\ \overline{\bf 10}_{\bf -1}+\big\{2({\bf 1}_{\bf -5} +{\bf
1}_{\bf 5}+  {\bf 5}_{\bf -3} + \overline{\bf 5}_{\bf 3})+
3(\overline{\bf 10}_{\bf -1} + {\bf 10}_{\bf 1}) \big\}
+22\{{\bf 1}_{\bf 0}\},\label{CT6}\\
&T_2\ :\ {\bf 1}_{\bf -5}+{\bf 5}_{\bf 3}
+11\{{\bf 1}_{\bf 0}\} + 4{\bf D} + {\bf O10} + {\bf O\overline{16}},
\label{CT2}\\
&T_4\ :\ {\bf {5}}_{\bf -2}+ 2({\bf {5}}_{\bf -2}+\overline{\bf
5}_{\bf 2})+30 \{{\bf 1}_{\bf 0}\} + 12 {\bf D},\label{CT4} \\
&T_1\ :\ 2(\overline{\bf 5}_{\bf -\frac12})+2({\bf 5}_{\bf
+\frac12}) +
6({\bf 1}_{\bf -\frac52})+6({\bf 1}_{\bf +\frac52})+
 ({\bf D}{\bf 1}_{\bf +\frac52}) + ({\bf D}{\bf 1}_{\bf
-\frac52}),
\label{ECT10}\\
&T_7\ :\ 2({\bf 5}_{\bf +\frac12})+2(\overline{\bf 5}_{\bf
-\frac12})
+6({\bf 1}_{\bf +\frac52})+ 6({\bf 1}_{\bf
-\frac52}) + ({\bf D}{\bf 1}_{\bf +\frac52}) + ({\bf D}{\bf 1}_{\bf
-\frac52}),
 \label{ECT50}
\end{align}
where ${\bf D}$ denotes a doublet of SU(2)$'$,
and ${\bf O10}$ and ${\bf O\overline{16}}$ mean ${\bf 10}$
and ${\bf \overline{16}}$ of SO(10)$'$, respectively. From the
hidden sector, there are twenty SU(2)$'$ doublets and one ${\bf
\overline{16}}'$ and one ${\bf 10}'$ of SO(10)$'$. Note that matter
fields from T1 sector and T7 sector have exotic representations
which are not present in the MSSM. We name them by G-exotics for
(anti)-fundamental representations of SU(5) and by E-exotics for
singlets of SU(5) but with nonvanishing U(1)$_X$ so that they
have exotic U(1)$_{\rm em}$ charge.

The superpotential terms  are obtained by examining vertex operators
satisfying the ${\bf Z}_{12-I}$ orbifold conditions \cite{Hamidi:1986vh,
ChoiKim06}. It can be summarized by the following selection rules:
\begin{itemize}
\item H-momentum conservation with $\phi_s =
  \left(\frac{5}{12}, \frac{4}{12}, \frac{1}{12} \right)$.
\begin{eqnarray}
\sum_Z R_1 (z) = -1 {\rm~ mod~} 12 , \quad
\sum_z R_2 (z) = 1 {\rm~ mod~} 3,\quad
\sum_z R_3 (z) = 1 {\rm~ mod~} 12,
\end{eqnarray}
where $z(\equiv A,B,C,\dots)$ denotes the index of states participating
in a vertex opeartor.
\item Space group selection rules:
\begin{eqnarray}
&& \sum_z k(z) = 0 {\rm~ mod~} 12, \\
&& \sum_z \left[ km_f \right] (z) = 0 {\rm~ mod~} 3.
\end{eqnarray}
\end{itemize}
Neglecting oscillator numbers, the $H$-momenta for ${\bf Z}_{12-I}$
twist are
\begin{equation}
\begin{split}
&U_1: \textstyle (-1~0~0),\quad  U_2: \textstyle (0~1~0),\quad
 U_3: \textstyle (0~0~1),\quad \\
&T_1: \textstyle (\frac{-7}{12}~\frac{4}{12}~\frac{1}{12}),\quad
T_2: (\frac{-1}{6}~\frac{4}{6}~\frac{1}{6}),\quad
 T_3: (\frac14~0~\frac{-3}{4})\\
&T_4:  \textstyle (\frac{-1}{3}~\frac13~\frac13),\quad T_5:
(\frac{1}{12}~\frac{-4}{12}~\frac{-7}{12}),\quad
 T_6:(\frac{-1}{2}~0~\frac12)
 \end{split}
 \label{Hmomenta}
\end{equation}

Now we present superpotential terms relevant for the low energy
phenomenology. Such terms include the mass terms for the exotics,
the $\mu$ terms and the Yukawa couplings. We summarize them in
Tables \ref{Superpotentials} and \ref{yukawas}.

\TABLE{
\caption{ Superpotential terms relevant for phenomenology: 1. Mass
terms for vector-like representations ($\mu$-like terms), G-exotic
fields and E-exotic fields, and the $\mu$ terms for  $H_u$ and $H_d$.
We represent superpotential terms up to dimension 5 only for mass terms
of E-exotic fields since there are too many terms from dimension 6.
}\label{Superpotentials}
\begin{tabular}{|rl||rl|}
\hline \multicolumn{4}{|c|}
{Mass terms for vector-like representations} \\
\hline \multicolumn{2}{|c|}{ $\left\{
\begin{array}{c}
T6_{10} T6_{\bar{10}} \\
T6_{5} T6_{\bar{5}} \\
T6_{1I} T6_{1J}
\end{array}
\right\} \times \left\{
\begin{array}{c}
C^0_5 C^0_6 C^-_3 T2^+_{O\overline{16}} T2^+_{O\overline{16}} T2^-_{O10}, \\
C^0_5 C^0_6 C^0_7 C^-_3 T2^-_{O10} T2^-_{O10}
\end{array}
\right\} $ } & \multicolumn{2}{|c|}{
 $T4_5 T4_{\bar{5}} \times
\left\{
\begin{array}{c}
s^0_4,~~~ s^u s^0_2, \\
s^0_4 (h_3 \bar{h}_1 + s^0_4 h_1 \bar{h}_3 ), \\
 s^u s^0_2 (h_3 \bar{h}_1 +  h_1 \bar{h}_3 ),  \\
C^0_6 s^0_2 s^0_4 ( C^0_6 h_3 + C^0_5 \bar{h}_1 )
\end{array}
\right\} $ }
\\
\hline\hline
\multicolumn{4}{|c|}{$\mu$ terms } \\
\hline \multicolumn{4}{|c|}{ $ T4_5 U2_{\bar{5}} \times \left\{
\begin{array}{c}
  s^0_2 s^0_4, \qquad s^u s^0_2 s^0_2, \qquad
  s^0_2 s^0_4 (h_3 \bar{h}_1 + h_1 \bar{h}_3 )\\
  s^u s^0_2 s^0_2 ( h_3 \bar{h}_1 + h_1 \bar{h}_3 ),
  D^+_1 D^-_1 s^0_4 ( h_1 \bar{h}_4 + h_4 \bar{h}_1 ),
  D^+_1 D^-_2 s^0_4 ( h_2 \bar{h}_3 + h_3 \bar{h}_2 ), \\
  C^0_3 s^0_3 (s^0_4 s^0_8 h_1 + d^+_3 d^-_1 h_1 + s^0_4 s^0_5 \bar{h}_1
+ d^+_2 d^-_1 \bar{h}_1 ), \\
  C^0_6 s^0_2 s^0_2 s^0_4 h_3 ,  C^0_3 s^0_1 s^0_4 s^0_8 h_3 ,
 C^0_3 s^0_1 d^+_3 d^-_1 h_3,  \\
  C^0_5 s^0_2 s^0_2 s^0_4 \bar{h}_1, C^0_3 s^0_1 s^0_4 s^0_5 \bar{h}_3,
 C^0_3 s^0_1 d^+_2 d^-_1 \bar{h}_3
\end{array}
\right\}
$}\\
\hline\hline \multicolumn{4}{|c|}{Mass terms for G-exotics ($T1_{5},
T1^-_{5}, T7_{5}, T7^-_{5},
  T1_{\bar{5}A}, T1_{\bar{5}B}, T7_{\bar{5}A}, T7_{\bar{5}B}$  ) }\\
\hline
$T1^-_5 T1_{\bar{5}A} \times $
 &  $s^+_1 \bar{h}_3  $ & $T1^-_5 T7_{\bar{5}A} \times$
& $s^+_1 ( 1 +  h_1 \bar{h}_3 + h_3 \bar{h}_1 ) $  \\
$T1^-_5 T7_{\bar{5}B} \times$
& $C^0_3 C^0_6 ( s^0_5 s^0_8 s^+_1 + s^+_2 s^+_3 s^-_1 )$ &
$T7_5 T1_{\bar{5}A} \times$ &  $s^0_2 (1 + h_3 \bar{h}_1 + h_1 \bar{h}_3 )$
\\
$T7^-_5 T1_{\bar{5}A} \times$& $s^+_1 ( 1 + h_3 \bar{h}_1 + h_1 \bar{h}_3 )$ &
$T7_5 T7_{\bar{5}A} \times$&  $(s^0_2 h_1 + C^0_3 s^0_1 s^0_5 + D^+_1 D^-_2 h_2)$
\\
$T7^-_5 T7_{\bar{5}A} \times$ &  $s^+_1 h_1$ &
$T7_5 T1_{\bar{5}B} \times$ & $ C^0_3 s^0_5 (s^0_8 h_3 + s^0_5 \bar{h}_3)$\\
$T7^-_5 T1_{\bar{5}B} \times$ & $ C^0_3 C^0_5 ( s^0_5 s^0_8 s^+_1+
 s^+_2 s^+_3 s^-_1 )$ &
$T7_5 T7_{\bar{5}B} \times$ & $C^0_3 \left(
\begin{array}{l}
s^0_8 s^0_8 h_1 h_1 + s^0_5 s^0_8 h_1 \bar{h}_1 \\
+ s^0_5 s^0_5 \bar{h}_1 \bar{h}_1
\end{array}
\right)$ \\
\hline
\hline
%
\multicolumn{4}{|c|}{Mass terms for E-exotics}
\\
\hline
$T1^+_{1C} T7_{1A} \times$  & $  d^-_1 ( 1+
 h_3 \bar{h}_1 + h_1 \bar{h}_3 )$
& $ T1_{1E}  T7^-_{1B}
\times$ & $ s^+_1 (1 + h_3 \bar{h}_1 +  h_1 \bar{h}_3 )$ \\
$T1_{1E} T7_{1B} \times$ & $ C^0_5 C^0_7$ &
$T1_{1B} T1_{1E} \times$ & $ (C^0_5 s^0_3 s^0_6
 +C^0_1 s^0_6 s^0_7 +C^0_5 C^0_7 \bar{h}_3 )$ \\
$T1_{1B} T1_{1D} \times$ & $( C^0_5 s^0_6 s^0_8 + C^0_5 s^+_3 s^-_2 )$ &
$T1^+_{1D} T7_{1A} \times$ & $(s^-_1 h_3 \bar{h}_2 + s^-_1 h_2 \bar{h}_3 )$ \\
$T7_{1B} T7_{1D} \times$ & $( C^0_6 s^0_5 s^0_7 + C^0_6 s^+_2 s^-_3 ) $ &
$T7_{1A} T7_{1G} \times$ & $ \bar{h}_4 ( C^0_2 C^0_7 +  C^+_2 C^-_3 ) $ \\
$T7_{1D} T7^+_{1B} \times$ & $C^0_6 s^0_5 s^-_3$ &
$T1_{1A} T7^+_{1C} \times$ & $ d^-_1 h_2 \bar{h}_2 $ \\
$T1_{1A} T7^+_{1D} \times$ & $ s^-_1 ( h_3 \bar{h}_2 + h_2 \bar{h}_3 )$ &
$T7_{1B} T7^+_{1D} \times$ & $
\left\{ \begin{array}{l}
C^0_2 s^0_7 s^-_2 + C^0_6 s^0_1 s^-_3 \\
+C^0_2 s^0_6 s^-_3 +C^0_2 s^-_2 s^-_3
\end{array} \right\}$ \\
\hline
\end{tabular}
}

\TABLE{
\caption{ \label{yukawas} Super potentential terms relevant to the
low energy phenomenology {\it continued}: 2. Yukawa couplings of
MSSM matter fields.}
\begin{tabular}{|c|c||c|c|}
\hline
\multicolumn{4}{|c|}{Yukawa Couplings} \\
\hline \hline
Generation & \multicolumn{3}{|c|}{ Matter } \\
\hline
1 & \multicolumn{3}{|c|} { $U1_{\overline{10}}$, $U1_5$, $U1_1$ } \\
2 & \multicolumn{3}{|c|} { $U3_{\overline{10}}$, $U3_5$, $U3_1$ } \\
3 & \multicolumn{3}{|c|} { $T6_{\overline{10}}$, $T2_5$, $T2_1$ } \\
\hline\hline
(i,j) & $\bar{\bf 10}_i \bar{\bf 10}_j \bar{\bf 5}_H$
&               (i,j) & $\bar{\bf 10}_i {\bf 5}_j {\bf 5}_H$ \\
\hline (3,3) & $T6_{\bar{10}} T6_{\bar{10}} U2_{\bar{5}}
\times \{ 1 , h_3 \bar{h}_1 , h_1 \bar{h}_3 \}$
&               (3,3) & $T6_{\bar{10}} T2_5 T4_5
\times \{ 1 , h_3 \bar{h}_1 , h_1 \bar{h}_3 \}$  \\
(3,2) & $T6_{\bar{10}} U3_{\bar{10}} U2_{\bar{5}} h_1$ &
(3,2) &  $T6_{\bar{10}} U3_5 T4_5 C_6^0 $ \\
(3,1) & $T6_{\bar{10}} U1_{\bar{10}} U2_{\bar{5}} \bar{h}_3$ &
(3,1) &  $T6_{\bar{10}} U1_5 T4_5 C_5^0 $ \\
(2,2) & $U3_{\bar{10}} U3_{\bar{10}} U2_{\bar{5}} h_1 h_1$ &
              (2,3) &  $U3_{\bar{10}} T2_5 T4_5 h_1$ \\
(2,1) & $U3_{\bar{10}} U1_{\bar{10}} U2_{\bar{5}} \times
\{ 1 , h_3 \bar{h}_1 , h_1 \bar{h}_3 \}$
&               (2,2) &  $U3_{\bar{10}} U3_5 T4_5 C^0_6 h_1$ \\
(1,1) &  $U1_{\bar{10}} U1_{\bar{10}} U2_{\bar{5}} \bar{h}_3
\bar{h}_3$  &
(2,1) & $U3_{\bar{10}} U1_5 T4_5 C_5^0 h_1$ \\
&     &         (1,3) & $U1_{\bar{10}} T2_5 T4_5 \bar{h}_3$ \\
&     &         (1,2) &
 $U1_{\bar{10}} U3_5 T4_5 C_6^0 \bar{h}_3$ \\
&     &         (1,1) & $U1_{\bar{10}} U1_5 T4_5 C^0_5 \bar{h}_3$ \\
\hline
(i,j) & ${\bf 5}_i {\bf 1}_j \bar{\bf 5}_H$ & & \\
\hline
(3,3) & $T2_5 T2_1 U2_{\bar{5}} s^0_2 s^0_2$ & & \\
(3,2) &  $T2_5 U3_1 U2_{\bar{5}} s^0_2 \bar{h}_1$  & & \\
(3,1) &  $T2_5 U1_1 U2_{\bar{5}} s^0_2 h_3$ & & \\
(2,3) &  $U3_5 T2_1 U2_{\bar{5}} s^0_2 \bar{h}_1 $  & & \\
(2,2) &  $U3_5 U3_1 U2_{\bar{5}} \bar{h}_1 \bar{h}_1$  & & \\
(2,1) & $U3_5 U1_1 U2_{\bar{5}} \times \{1, h_3 \bar{h}_1 , h_1
\bar{h}_3 \}$ & & \\
(1,3) &  $U1_5 T2_1 U2_{\bar{5}} s^0_2 h_3$ & & \\
(1,2) & $U1_5 U3_1 U2_{\bar{5}} \times \{1, h_3 \bar{h}_1, h_1
\bar{h}_3\} $ & & \\
(1,1) &  $U1_5 U1_1 U2_{\bar{5}} h_3 h_3$  & & \\
\hline
 \end{tabular}
}

\subsection{Hidden sector SO(10)$^\prime$ and
SU(2)$^\prime$}\label{subsec:hiddensector}

There are two hidden sector nonabelian groups,  SO(10)$^\prime$ and
SU(2)$^\prime$. Inspecting SU(2)$^\prime$ matter spectrum, there are
twenty SU(2)$^\prime$ doublets: $\delta_1$, $\delta_2$, $\delta_3$,
$\delta_4$,  $D_1^+$, $D_1^-$, $D_2^+$, $D_2^-$, $d_1^+$, $d_1^-$,
$d_2^+$, $d_2^-$, $d_3^+,$ and $d_3^-$. Inspecting the
SO(10)$^{\prime}$ matter representation, we find only a single
SO(10)$^\prime$ spinor $\overline{\bf 16}^\prime$ and a single
SO(10)$^\prime$ vector $\bf 10^\prime$. Since there are four
nonabelian gauge groups in total after the flipped SU(5) breaking,
it seems that we need to consider four $\theta$
parameters, those of SU(3)$_c$, SU(2)$_W$, SO(10)$^\prime$ and
SU(2)$^\prime$. In fact, among those, we need to consider only SU(3)$_c$
and SO(10)$^\prime$ because SU(2)$_W$ is broken at the electroweak scale
and we assume that SU(2)$^\prime$ is broken.  For SO(10)$^\prime$,
here we just assume that a subgroup of SO(10)$^\prime$ confines at an
intermediate scale since the hidden sector dynamics is not well
understood yet at present.

\section{Approximate \UPQ\  Symmetry of $\Z_{12-I}$ Model }
\label{sec:u1s}

In this section, we find \UPQ\ symmetries allowed from the
superpotential terms in the $\Z_{12-I}$ model. To all orders, only
gauge and discrete symmetries can remain valid since string theory
{\it a priori} does not impose any continuous global symmetry.
However, at a certain order of the effective Lagrangian, the theory
can have approximate global symmetries. Such approximate symmetries
are not given from the first principle, and thus we have to
enumerate allowed symmetries by reading all superpotential terms. To
do this, we have made a computer program which enumerate all the
allowed potential terms and continuous abelian symmetries at a given
order.

If we assign U(1) charge $q_j$ for each field $\Phi_j$, a
superpotential term $W_i \propto \prod_j (\Phi_j)^{\lambda_{ij}}$
has U(1) charge $\sum_j \lambda_{ij} q_j$, where $\lambda_{ij}$
represents the number of occurrences of the field $\Phi_j$ in the
$i$-th superpotential term. Thus, the algorithm for finding out U(1)
charges is basically the same as finding eigenvectors with zero
eigenvalue of the matrix $\lambda^T \lambda$. It is remarkable that
we can find the eigenvectors exactly (without any numerical error)
since the eigenvectors can always be represented by integer-valued
vector due to the fact that the matrix $\lambda^T \lambda$ has only
integer-valued components and we are only interested in the
eigenvectors with zero-eigenvalue. Eigenvectors with zero eigenvalue
can be obtained by the Gauss elimination method: By multiplying
\begin{eqnarray}
\left(\begin{array}{cccccc}
1 & & & & & \\
 & \ddots & & & & \\
 & & 1 & & c &  \\
 & & & \ddots & & \\
 & & & & &  \\
 & & & & & 1
\end{array} \right)
\mbox{ or }
\left(\begin{array}{cccccc}
 1 & & & & & \\
 & \ddots & & & & \\
 &  & 0 &  & 1 & \\
 &  & & \ddots & & \\
 & & 1 & & 0  & \\
 & & & & & \ddots
\end{array} \right)
\end{eqnarray}
onto $\lambda^T \lambda$, the eigenvector solutions are not changed.
Using such operations, one can reduce the matrix up to the following
upper triangular form
\footnote{Note that we cannot make the matrix be the identity since
the matrix is not invertible. }
\begin{eqnarray}
\left(
\begin{array}{cccc}
a_{11} & a_{12} & \cdots & a_{1N} \\
0 & \ddots & \cdots & \vdots \\
\vdots &  & \ddots  & \\
0 & \cdots &0 & a_{NN}
\end{array}
\right),
\end{eqnarray}
where  $N$ is the total number of field species,
the diagonal element $a_{ii}$ is either 1 or 0,
$a_{ij} = 0 $ when $a_{ii} = 0$ and $a_{ji} = 0 $ when $a_{ii} = 1$.
Denote nonzero element of $a_{ij}$ by $b_{r \alpha}$ where
$r$ runs over $k$ indices of nonzero diagonal elements
and $\alpha$ runs over $N-k$ indices of zero diagonal elements.
Then, the eigenvectors are
\begin{eqnarray}
\{q_r | q_\alpha\} &=& \{ ( -b_{r 1}) \quad~~| 1 0 0 \dots 0 \} \\
\{q_r | q_\alpha\} &=& \{ (-b_{r 2})  \quad~~| 0 1 0 \dots 0 \} \\
\vdots & & \vdots \\
\{q_r | q_\alpha\} &=& \{ (-b_{r (N-k)} ) | 0 \dots  1 \}  ,
\end{eqnarray}
where $q_r$ are the eigenvector components of the $r$-th element
with nonzero diagonal element for the above matrix, and $q_\alpha$
are those of zero diagonal element.

We have found all the superpotential terms up to dimension 9.
From these, we have found U(1) symmetries using the algorithm described
above. Certainly, six of them are U(1) gauge symmetries.
Obtaining all the gauge anomalies, we rediagonalize the gauge symmetries
according to the anomalies:
\begin{eqnarray}
&&\left\{\begin{array}{l}
Q_X= -Z_1\\
Q_1= Z_2+6Z_3\\
Q_2= -Z_2 + 6 Z_4\\
Q_3= Z_5 \\
Q_4= 2 Z_2+ 3Z_6
\end{array}\right.\label{anmfree}\\
&&\quad\begin{array}{l} Q_{\rm an}= -6Z_2+Z_3-Z_4+4Z_6
\end{array}\label{anmlous}
\end{eqnarray}

Five U(1)s of (\ref{anmfree}), U(1)$_X$,U(1)$_1,\cdots,$ and
U(1)$_4$, do not carry any gauge and gravitational anomalies while
the \Uan\ of (\ref{anmlous}) is anomalous.
Therefore, this model has an anomalous gauge U(1) which can be
consistent only with Green-Schwarz mechanism \cite{Green:1984sg,Barr:1985hk}.
Note that the anomaly check is very nontrivial, and hence this
gaurantees the matter spectrum shown in Table \ref{spectrum} and
\ref{spectrum2} is correct.

Up to dimension 7, we find one global U(1) symmetry which does not
belong to the above gauge symmetries.
 This accidental global symmetry must be broken
at still higher order superpotential terms. We discuss it in
Subsec. \ref{approxU1}

The anomalous gauge U(1) induces the Fayet-Illiopoulos term.
D-flat condition for the anomalous \Uan\ is required to be
$$
\langle D^{({\rm an})}\rangle=\left\langle\frac{2g}{192\pi^2}{\rm
Tr}X_{\rm an}+\sum_i X_{\rm an}(i)\phi^*(i)\phi(i)\right\rangle=0.
$$
Some singlet fields get VEVs to satisfy the above equation, breaking
some U(1) symmetries. But we will not focus on the details of this
mechanism. Below, we try to find a vacuum (or vacua) realizing
observable QCD axion. For this objective, certainly we will choose
some VEVs of singlet scalar fields. In doing so, we will restrict
such that all the known low energy phenomena such as fermion masses
are reproduced. So, we will consider \Qem$=\frac23, -\frac13$ quarks
and charged leptons. Neutrino masses are not considered here since
the mass matrix is too gigantic because of the appearance of
numerous singlets.

\TABLE{
\caption{ \label{tab:Uan} Non-vanishing \Uan\ charges. Fields
$\Phi({\cal P})$ denotes the multiplicity ${\cal P}$ in case ${\cal
P}>1$.}
\begin{tabular}{|cc|cc|cc|cc|cc|}
\hline
 {fields}& $Q_{\rm an}$&{fields}& $Q_{\rm an}$&{fields}&
$Q_{\rm an}$&{fields}& $Q_{\rm an}$&{fields}& $Q_{\rm an}$ \\
\hline
 $U3_{5}$           &  --3  & $U3_{\overline{10}}$ &  --3 &
 $U3_1$             &  --3  & $U2_{\bar{5}}$      &    6 &
 $U1_{\overline{10}}$ &  --3  \\
 $U1_{\bf5}$         &  --3  & $U1_1$             &  --3 &
 $T1_{\bar{5}A} $    &    4  & $T1_{\bar{5}B}$      & --3  &
 $T1_{1A}$          &    3  \\
 $T1_{1B} $         &  --2  & $T1_{1C} $          & --2  &
 $T1_{1D} $         &  --6  & $T1_{1E}$           &   1  &
 $T1_{1F}$          &    1  \\
 $T1_{1G}$          &    1  & $\delta_1$         &  --4  &
 $\delta_2$         &  --1  & $T1^+_{1D}$        &    3  &
 $T1^-_{5}$         & --2   \\
 $T1^-_{1A}$        & --6   & $T1^-_{1B}$        &    1  &
 $T1^-_{1C}$        & --3   &  $T2_{5}$          &    1  &
 $T2_{1}$           &   1   \\
 $T2^+_{O\overline{16}}$ & --3 & $T2^-_{O10}$         & --3 &
 $C^0_1$           & --1   & $C^0_2$             & --1  &
 $C^0_3$           & --3   \\
 $C^0_4$             & --3  & $C^0_5$           &   4  &
 $C^0_6$           &   4   & $C^0_7$             & --3  &
 $D^-_1 $          & --1   \\
 $D^-_2 $            & --1  & $C^-_1$           & --5  &
 $C^-_2$           & --5   &  $C^-_3$            &   1  &
$D^+_1 $            & --3  \\
 $D^+_2 $          & --3   & $C^+_1$             &   1  &
 $C^+_2$           &   1   & $T4_{\bf 5}(3)$    &   2   &
 $T4_{\bar{\bf 5}} (2)$ &  2 \\
 $s^0_1 (2)$       &  --4  & $s^0_2 (2)$         & --4   &
 $s^0_3 (2)$       & --4   & $s^0_4 (2)$         & --4   &
 $s^0_5 (2)$       &   3   \\
 $s^0_6 (2)$       &   1   & $s^0_7 (2)$         &   1   &
 $s^0_8 (2)$       &   3   & $d^+_1 (2)$         & --6   &
 $s^+_1 (3)$       &  -2   \\
 $d^+_2 (2)$       &   1   & $s^+_2 (2)$         &   5   &
 $d^+_3 (2)$       &   1   & $s^+_3 (2)$         &   5   &
 $d^-_1(2)$        & --2   \\
 $s^-_1 (3)$       & --6   & $d^-_2(2)$          &   3  &
 $s^-_2 (2)$       & --1   & $d^-_3(2)$          &   3  &
 $s^-_3 (2)$       & --1   \\
 $T6_{\bf 5}(2)$   & --3   & $T6_{\overline{\bf 10}}(4)$ &  --3 &
 $T6_{1I}(2)$       & --3  & $T6_{\bar{\bf 5}}(2)$   &   3 &
 $T6_{\bf 10}(3)$    &   3  \\
 $T6_{1J}(2)$       &   3  & $T7_{\bar{\bf 5}A}$     &   4 &
 $T7_{\bar{\bf 5}B}$ & --3  & $T7_{1A}$              &   3 &
 $T7_{1B}$          & --2  \\
 $T7_{1C}$          & --2 &  $T7_{1D}$              & --6 &
 $T7_{1E}$          &   1 &  $T7_{1F}$              &   1 &
 $T7_{1G}$          &   1 \\
 $\delta_3$         & --4 & $\delta_4$             & --1 &
 $T7^+_{1D}$        &   3 & $T7^-_{\bf 5}$           & --2 &
 $T7^-_{1A}$        & --6 \\
 $T7^-_{1B}$        &   1 & $T7^-_{1C}$             & --3 &
                   &     &                     &      &
                    &    \\
\hline
\end{tabular}
}

Low energy fields and the anomalous gauge U(1)$_{\rm an}$ charges
are listed in Table \ref{tab:Uan}.
 $D,d,$ and $\delta$ fields are
SU(2)$^\prime$ doublets. One can easily check that the
\Uan--SU(5)--SU(5), \Uan--SU(2)$'$--SU(2)$'$, and
\Uan--SO(10)$'$--SO(10)$'$ anomalies are {universal}, which is
required by Green-Schwarz mechanism.

We must assign some singlets GUT scale VEVs for successful Yukawa
couplings.  First, the singlet fields combinations which appear in
the Yukawa couplings must have GUT scale VEVs generically. Since the
exotic fields which appear in  $T1$ and $T7$ sectors must have the
GUT scale mass not to spoil the gauge coupling unification, the
singlet fields in those mass term must be of the order the GUT scale
also.

\subsection{Exotics}

There are two kinds of exotics, G-exotics and E-exotics. They are
named according to SU(5): G-exotics are SU(5)$_X$ (anti-)quintets
and E-exotics are SU(5) singlets (with nonvanishing $X$ charges).

We will show shortly in Subsec \ref{SMYukawa} that $s_2^0$ {should
be} small to explain the electron mass successfully. Under this
circumstance, we try to remove exotics at the GUT scale.

\subsubsection{G-exotics}
 We have 4 pairs of G-exotics from T1 and T7
sectors:
\begin{align}
 &{\bf 5}_{1/2}:T1_5 , T1^+_5, T7_5, T7^-_5\nonumber\\
 & {\overline{\bf 5}}_{-1/2}: T1_{\bar{5}A},
T1_{\bar{5}B}, T7_{\bar{5}A}, T7_{\bar{5}B}\nonumber
 \end{align}
In terms of SM gauge group $SU(3)_C \times SU(2)_L \times U(1)_Y$,
they transform as ${\bf 5}_{1/2} \rightarrow ({\bf  \overline{3}, 1
})_{\frac{1}{6}} + ({\bf 1, 2 })_0 $ and ${\bf \overline{5}}_{-1/2}
\rightarrow ({\bf 3,1})_{-\frac{1}{6}} + ({\bf 1,2})_0 $. In Table
\ref{Superpotentials}, we show the mass terms for G-exotics up to
dimension 7. Some mass terms at dimension 6 and 7 are omitted if
there are lower dimensional terms of the same type since they are
too many. All those mass terms which do not appear in the table have
terms at dimension 8 except for $T1_5 T1_{\bar{5}B}$ and $T1_5 T7_{\bar{5}B}$.

\subsubsection{E-exotics}

There are sixteen E-exotics. It is not wieldy to calculate the
determinant of the E-exotics mass matrix. Therefore, we first find
out approximate U(1) directions from other Yukawa couplings and then
look for the E-exotics mass matrix.

\subsection{Vectorlike pairs}

Another consideration is the mass terms for the following
vector-like pairs,
\begin{equation}
(T6_5, T6_{\bar{5}}), ( T6_{10} , T6_{\overline{10}} ), ( T6_{1I},
T6_{1J}), \quad (T4_5, T4_{\bar{5}}).\label{VecLike}
\end{equation}
The vector-like pairs in $T6$-sector of (\ref{VecLike}) are exactly
vector-like under all gauge symmetries. Also, they have exactly the
same $H$-momenta, which means that all the vector-like pairs $(T6_5,
T6_{\bar{5}}), ( T6_{10} , T6_{\bar{10}} ), ( T6_{1I}, T6_{1J})$
satisfy the same selection rules for the superpotential.

To break the flipped SU(5) down to the SM gauge group, vectorlike
pairs ${\bf 10}+\overline{\bf 10}$ of T6 must develop VEVs. Here,
note that the multiplicities are different: ${\cal P}(\overline{\bf
10})=4$ and ${\cal P}({\bf 10})=3$. One $\overline{\bf 10}$ is not
matched and it becomes the third family member called $\overline{\bf
10}_t$ of the SM. Thus, the coupling allowed by $T6T6$ is restricted
to three vectorlike pairs of ${\bf 10}+\overline{\bf 10}$, and
$\overline{\bf 10}_t$ carries an independent phase.  At the lowest
order, the resultant mass terms for three vectorlike pairs are
presented in Table \ref{Superpotentials}. Those two terms involve
the hidden sector matter fields $T2^+_{O\overline{16}}$ and
$T2^-_{O10}$. $T2^+_{O\overline{16}}$ cannot have a VEV at the GUT
scale since it leads to the D-term SUSY breaking at that scale. A
VEV of $T2^-_{O10}$ has no problem with SUSY breaking, but here we
do not assume about the symmetry breaking pattern in the hidden
sector. At dimension 9, which is next to the lowest order, there are
many terms inducing mass terms for the vectorlike pairs. In this
paper, we do not show all the dimension 9 terms since there are too
many of them. Since we will argue that $C^0_5$ and $C^0_6$ must have
a VEV near the GUT scale as shown in Subsec. \ref{SMYukawa}, we show
the mass terms involving $C^0_5$ and $C^0_6$:
\begin{align}
\left\{
\begin{array}{c}
T6_{10} T6_{\bar{10}} \\
T6_{5} T6_{\bar{5}} \\
T6_{1I} T6_{1J}
\end{array}
\right\} \times C^0_3 C^0_5 C^0_6 s^-_1
\left\{
\begin{array}{c}
s^+_3 s^0_1 h_3 + s^+_2 s^0_1 \bar{h}_3 \\
+  s^+_3 s^0_3 h_1 + s^+_2 s^0_3 \bar{h}_1
\end{array}
\right\}.
\end{align}
Therefore, if a combination of ($C^0_3$, $s^-_1$, $s^+_2$ or
$s^+_3$, $s^0_1$ or $s^0_3$) has a ${\cal{O}}(M_{\rm GUT})$ scale
VEV, the vectorlike pairs can have ${\cal{O}}(M_{\rm GUT})$ scale
masses.

\subsection{Approximate global symmetry U(1)$_{\rm glA}$}
\label{approxU1}

Before proceeding to discuss the global symmetry we have found
and used it as a PQ symmetry for QCD axion, we first discuss
the difficulty of U(1)$_{\rm PQ}$ with the intermediate PQ symmetry
breaking scale around $10^{10-12}$ GeV in $Z_{12-I}$ orbifold flipped
SU(5) model.

The following fields are non-singlet under SM gauge group.
MSSM matter and Higgs fields :
\begin{align}
\begin{array}{ccc}
U1_{\overline{10}}, & U1_{5}, & U1_{1}, \\
U3_{\overline{10}}, & U3_{5}, & U3_{1}, \\
T6_{\overline{10}}, & T2_{5}, & T2_{1},
\end{array} \\
T4_5, U2_{\bar{5}},
\end{align}
Vector-like pairs of SM matter-like representation :
\begin{align}
& ( T4_5, T4_{\bar{5}} ), \nonumber \\
& (T6_{10}, T6_{\overline{10}}), \nonumber\\
& (T6_{5}, T6_{\bar{5}}), \nonumber \\
& (T6_{1I}, T6_{1J}), \nonumber
\end{align}
and vector-like pairs of G-exotics and E-exotics from the
1st and the 7th twisted sectors.

One pair of $T6_{10}$ and $T6_{\overline{10}}$ must have GUT scale
VEV to break the flipped SU(5) gauge group to the SM ones and then
$T6_{5}, T6_{\bar{5}}$ and $T6_{1I} T6_{1J}$ must also have the mass
of ${\cal O}(M_{\rm GUT})$ {not to spoil gauge coupling unification.
} { On the other hand, $T4_5$ and $T4_{\bar{5}}$ do not need to have
the GUT scale mass since they form a complete SO(10) multiplet. }
Note that the nonzero PQ charges of $T6_{10}$ and
$T6_{\overline{10}}$ do not necessarily mean PQ symmetry is broken
at the VEV scale of those fields because the Goldstone boson of the
broken symmetry is eaten as a longitudinal degree of freedom of the
massive gauge boson and one linear combination of the original
global symmetry and one of $SU(5) \times U(1)_X$ can remain as a
surviving low energy global symmetry, which we called the 't Hooft
mechanism \cite{'tHooft:1971rn}. However, to have such linear
combination exist, $T6_{10}$ and $T6_{\overline{10}}$ must have a
vector-like pair of charges under \UPQ\ symmetry. { Furthermore, the
other fields which become massive at the GUT scale must have
vector-like charges since we want to break \UPQ\  at much lower
scale.}

We may try \UPQ\ symmetry to be flavor-dependent. However, then the
flavor-dependence is severely restricted if we consider the PQ
breaking symmetry scale around ${\cal O}(10^{12}\ {\rm GeV})$ since
the maximum hierarchy we have in the low energy Yukawa coupling is
$10^{-6}$ for the \Qem $=\frac23$ quarks.

Firstly, let us consider the flavor-independent PQ symmetry case.
Denote \UPQ\ charges of SM non-singlet fields by
\begin{align}
\begin{array}{l}
Q_{\overline{10}} = Q ( U1_{\overline{10}}) = Q(U3_{\overline{10}})
= Q(T6_{\overline{10}}), \\
Q_{5}
= Q ( U1_{5}) = Q(U3_{5}) = Q(T2_{5}), \\
Q_{1}
= Q( U1_1 ) = Q( U3_1 ) = Q(T2_1),  \\
Q_{5H} = Q( T4_{5} ),  \\
Q_{\bar{5} H} = Q(U2_{\bar{5}})
\end{array}   \label{u1classeq1}
\end{align}
The cubic Yukawa couplings $\overline{\bf 10}_i \overline{\bf
10}_j\bar{\bf 5}_H$, $\overline{\bf 10}_i {\bf 5}_j {\bf 5}_H$ and
${\bf 5}_i{\bf 1}_j \bar{\bf 5}_H $ are required to be invariant
under the symmetry we try to find, and we obtain the following
relations,
\begin{align}
&Q_{\bar{5} H } = - 2 Q_{\overline{10}}, \nonumber\\
&Q_{5 H} = - Q_{\overline{10}} - Q_5, \nonumber\\
&Q_1 = 2 Q_{\overline{10}} - Q_5.  \label{u1classeq2}
\end{align}
{ In addition to the SM Yukawa coupling, this model has another
cubic coupling involving SM non-singlets:
\begin{eqnarray}
&& T6_5 T4_{\bar{5}} T2_1 + T6_5 U2_{\bar{5}} T6_{1I} +
 T2_5 T4_{\bar{5}} T6_{1I}.
\end{eqnarray}
From the above couplings, we further get
\begin{align}
&Q(T4_{\bar{5}}) = Q_{\bar{5} H} = - Q_{\overline{10}},  \nonumber\\
&Q(T6_{5}) = -Q(T6_{\bar{5}}) = Q_5, \nonumber \\
&Q(T6_{1I}) = -Q(T6_{1J}) = Q_1 = 2 Q_{\overline{10}} - Q_5.
\label{u1classeq3}
\end{align}
Since this determines all the \UPQ charges for SU(5) nonsinglets,
we can calculate the \UPQ-SU(5)-SU(5) anomaly
in terms of $Q_{\overline{10}}$
and $Q_5$. One can easily check that it is just zero, and thus
we cannot have any approximate flavor-independent PQ symmetry
which has the \UPQ-QCD-QCD anomaly.}

This leads us to the flavor-dependent case. However, the
situation does not improve very much in this case either. Since the
down type Yukawa coupling $\overline{\bf 10}_i\overline{\bf
10}_j\bar{\bf 5}_H$ does not reveal large hierarchy among the
families, we have to assign the flavor-independent symmetry for the
field which take part in these Yukawa couplings. Thus, it still
holds that
\begin{eqnarray}
Q_{\overline{10}}
= Q ( U1_{\overline{10}}) = Q(U3_{\overline{10}})
= Q(T6_{\overline{10}}).
\end{eqnarray}
Additionally, trilinear couplings give a strong restriction in
choosing a PQ symmetry.  If a cubic combination of SM nonsinglet
field is not invariant and that combination appears with other
singlet fields in the superpotential, one can assign the
compensating U(1) charges for the involved singlet fields so that
the overall term is still invariant. However, for the trilinear
couplings of the SM fields which does not involve any singlet
fields, this cannot be done. Therefore, such terms must be required
to be invariant first. In the SM Yukawa coupling combination, we
have
\begin{eqnarray}
&& T6_{\overline{10}} T6_{\overline{10}} U2_{\bar{5}}
+ U3_{\overline{10}} U1_{\overline{10}} U2_{\bar{5}}
+ T6_{\overline{10}} T2_{5} T4_{5}  \nonumber \\
&& + U3_{5} U1_1 U2_{\bar{5}}
+ U1_5 U3_1 U2_{\bar{5}}.
\end{eqnarray}

Another consideration we should take care of is the overall mass
hierarchy of the SM matter fields. From the observed masses of
quarks and leptons, we conclude that the determinant of up-type
quark mass matrix and lepton mass matrix must have at most one
factor of $\frac{F_{PQ}}{M_{GUT}}$. This requires at least one of
22-component and 33-component of the up-type quark Yukawa couplings
must have no \UPQ\ charge. In fact, which component is
\UPQ-noninvariant is just a matter of the choice:
$U1_{\overline{10}}$ or $U3_{\overline{10}}$ is the lightest
spectrum ($u$-quark). Here, we discuss only the case where
$U1_{\overline{10}}$ is the lightest one. The resultant \UPQ\
concerning the above constraints is \footnote{Here, we assign
$Q_{\overline{10}} =1 $ without loss of generality and $Q^1_{5} = 0
$ using $U(1)_X$ gauge symmetry. }
\begin{align}
& Q_{\overline{10}} = 1, \nonumber \\
& Q^3_{5} = Q^2_{5} = q, \quad  Q^1_{5} = 0, \nonumber \\
& Q^3_{1} = Q^2_{1} = 2, \quad Q^1_{1} = 2 -q, \nonumber \\
& Q_{5H} = -1-q , \quad Q_{\bar{5} H} = -2, \nonumber
\end{align}
where the superscript means the generation number.
However, this assignment is not successful. For example,
the lowest order term of the 11-component of lepton Yukawa coupling is
$(s^0_2)^2 U1_{5} U1_{1} U2_{\bar{5}}$, and thus $Q(s^0_2)= -q/2$.
Determinant of lepton mass matrix is proportional to $(s^0_2)^2$, and so
$\langle s^0_2 \rangle$ is $10^{-3} M_{st}$ to fit the phenomenological
values. Therefore, PQ symmetry breaking scale is around $10^{15}$ GeV in
this assignment.

Up to now, we discussed the difficulty of obtaining the intermediate
scale ($\sim 10^{11}$ GeV) for the PQ symmetry breaking scale.
We must {\it resort to the GUT scale PQ symmetry
breaking scale.} For cosmological application, we may resort to the
anthropic principle, {\it a la} Ref. \cite{WilAnth}.
Then, not worrying about the PQ symmetry breaking scale, we look for
approximate global symmetries surviving up to a high order.

\TABLE{
\caption{ \label{tab:UglA} Non-vanishing U(1)$_{\rm glA}$ charges.
Fields $\Phi({\cal P})$ denotes the multiplicity ${\cal P}$ in case
${\cal P}>1$.}
\begin{tabular}{|cc|cc|cc|cc|cc|}
\hline
 {fields}& $Q_{\rm glA}$&{fields}& $Q_{\rm glA}$&{fields}&
$Q_{\rm glA}$&{fields}& $Q_{\rm glA}$&{fields}& $Q_{\rm glA}$\\
\hline
$U3_{5}$ &  4 & $U1_{5}$ & 4 &
$T2_{5}$ & 4  & $T4_{5}$(3) & --4 &
$T6_{\bar{5}}(2)$ & --4 \\
$T6_{5}(2)$ & 4 & $U3_1$ & --4 &
$U1_1$ & --4 & $T2_{1}$ & --4 &
$T6_{1I}(2)$ & --4 \\
$T6_{1J}(2)$ &  4  & $T2^-_{O10}$ &  2 &
$T2^+_{O\overline{16}}$ &  1  & $s^u$ &  4 &
$s^0_4 (2)$ &   4  \\
$s^0_6 (2)$ & --4  & $s^0_7 (2)$ & --4  &
$d^-_1(2)$  &   4  & $s^-_2(2)$  & --4  &
$s^-_3(2)$  & --4  \\
$C^0_1$     &   4  & $C^0_2$     &   4  &
$C^0_7$     & --4  & $C^+_1$     &   4  &
$C^+_2$     &   4  \\
$T1_{1A}$   & --4  & $T1_{1D}$    &  4  &
$T1_{1E}$   &   4  & $T1_{{1F}}$  &  4  &
$T1_{1G}$   &   4  \\
$\delta_1$ & --4  & $T1^+_{1D}$  &  4  &
$T1^-_{1A}$ & --4  & $T1^-_{1B}$  & --4 &
$T7_{1A}$   & --4 \\
$T7_{1D}$   &   4  & $T7_{1E}$    & 4  &
$T7_{1F}$   & 4    & $T7_{1G}$    & 4  &
$\delta_3$  & --4 \\
$T7^+_{1D}$ &   4  & $T7^-_{1A}$  & --4 &
$T7^-_{1B}$ & --4  &             &     &
           &      \\
\hline
\end{tabular}
}

Up to $D=7$ superpotential terms, we have a global symmetry which we
call U(1)$_{\rm glA}$. {This symmetry is flavor-independent, and so
it belongs to the class described in Eqs. (\ref{u1classeq1}),
(\ref{u1classeq2}) and (\ref{u1classeq3}). As explained, it does not
have U(1)$_{\rm glA}$-SU(5)-SU(5) anomaly.}
 Low energy fields and their U(1)$_{\rm glA}$
global charges are listed in Table \ref{tab:UglA}.  From Table
\ref{tab:Uan} and \ref{tab:UglA}, the trace of the nonabelian
anomalies are given by
\begin{align}\begin{array}{l}
{\rm Tr}\, \left( Q_{\rm an} \, T_{\rm SU(5)} \, T_{\rm SU(5)}\right)
\,=\, -9,   \\
{\rm Tr}\, \left(Q_{\rm an} \, T_{\rm SO(10)'} \, T_{\rm SO(10)'}\right)
\,=\, -9,  \\
{\rm Tr}\, \left( Q_{\rm glA}\,  T_{\rm SU(5)} \, T_{\rm SU(5)} \right)
\,=\, 0,  \\
{\rm Tr}\, \left( Q_{\rm glA}\, T_{\rm SO(10)'}\, T_{\rm SO(10)'}\right)
\,=\, 4 \end{array}
\label{nonabAn}
\end{align}
We find that there is a gauge symmetry U(1)$_{\rm glV}$ which is
very close to U(1)$_{\rm glA}$,
$$
Q_{\rm glV} = Z_1 + \frac16 Z_2 + \frac16 Z_3 -\frac16 Z_4
+ \frac16 Z_5 + \frac16 Z_6.
$$
The U(1)$_{\rm glV}$ charges are given in (\ref{nonabAn}) except one
field $T2^+_{O\overline{16}}$ for which  $Q_{\rm
glA}(T2^+_{O\overline{16}}) = -1$.

At the $D=8$ superpotential terms, U(1)$_{\rm glA}$ is broken.  The
following superpotential terms at dimension 8 break  U(1)$_{\rm
glA}$,
\begin{align}
 \Delta W = F({\rm other~fields})T2^+_{O\overline{16}}
T2^+_{O\overline{16}} T2^-_{O10},
\end{align}
where
\begin{align}
 F_1 =\ &
T6_{10} T2_5 T4_5 T4_5 C^-_2 +
T6_{10} T6_{\overline{10}} T4_5 U2_{\bar{5}} C^-_2 +
T4_5 T6_5 U2_{\bar{5}} T6_{\bar{5}} C^-_2  \nonumber \\
&+ T2_5 T4_5 T4_{\bar{5}} T6_{\bar{5}} C^-_2
 + T6_{\overline{10}} U2_{\bar{5}} T4_{\bar{5}} T6_{\bar{5}} C^-_2
 + T6_{10} T4_{\bar{5}} T4_{\bar{5}} T2_1 C^-_2  \nonumber \\
&+ T6_{10} U2_{\bar{5}} T4_{\bar{5}} C^-_2 T6_{1I}
 + T6_{10} T6_{\overline{10}} C^0_5 C^0_6 C^-_3
 + T2_5 T6_{\bar{5}} C^-_2 s^0_5 s^0_6    \nonumber \\
&+ T2_5 T6_{\bar{5}} C^-_2 s^0_7 s^0_8
 + T2_5 T6_{\bar{5}} C^-_2 s^+_2 s^-_2
 + T2_5 T6_{\bar{5}} C^-_2 s^+_3 s^-_3   \nonumber \\
&+ T4_5 T4_{\bar{5}} T2_1 C^-_2 T6_{1J}
 + T4_5 U2_{\bar{5}} C^-_2 T6_{1I} T6_{1J}
 + T4_5 U2_{\bar{5}} C^-_2 h_2 \bar{h}_1 \nonumber \\
&+ T4_5 U2_{\bar{5}} C^-_2 h_1 \bar{h}_2
 + T4_5 U2_{\bar{5}} C^-_2 h_4 \bar{h}_3
 + T4_5 U2_{\bar{5}} C^-_2 h_3 \bar{h}_4 \nonumber \\
&+ T6_5 T6_{\bar{5}} C^0_5 C^0_6 C^-_3
 + T2_5 T6_{\bar{5}} C^0_6 h_3   C^-_3
 + T2_5 T6_{\bar{5}} C^0_5 \bar{h}_1   C^-_3  \nonumber \\
&+ T2_1 C^-_2 s^0_5 s^0_6 T6_{1J}
 + T2_1 C^-_2 s^0_7 s^0_8 T6_{1J}
 + T2_1 C^-_2 s^+_2 s^-_2 T6_{1J}  \nonumber  \\
&+ T2_1 C^-_2 s^+_3 s^-_3 T6_{1J}
 + C^0_5 C^0_6 T6_{1I} T6_{1J}   C^-_3
 + T2_1 C^0_6 T6_{1J} h_3   C^-_3 \nonumber \\
&+ T2_1 C^0_5 T6_{1J} \bar{h}_1   C^-_3
\end{align}
\begin{align}
F_2 = \ &
T4_5 T4_{\bar{5}} C^-_2 d^+_2 d^-_2 +
+ T4_5 T4_{\bar{5}} C^-_2 d^+_3 d^-_3
+ T4_5 T4_{\bar{5}} C^0_6 C^-_2 h_2
 + T4_5 T4_{\bar{5}} C^0_5 C^-_2 \bar{h}_4 \nonumber \\
&+C^-_2 s^0_6 d^+_2 s^+_2 d^+_3 +
C^-_2 s^0_7 s^+_2 d^+_3 d^+_3  +
C^-_2 s^0_6 d^+_2 d^+_2 s^+_3   +
C^-_2 s^0_7 d^+_2 d^+_3 s^+_3   \nonumber \\
& + C^-_2 s^0_5 s^0_6 d^+_2 d^-_2  +
C^-_2 s^0_7 s^0_8 d^+_2 d^-_2  +
C^-_2 s^0_5 s^0_7 d^+_3 d^-_2  +
C^-_2 d^+_2 s^+_2 d^-_2 s^-_2  \nonumber \\
& + C^-_2 s^0_6 s^0_8 d^+_2 d^-_3  +
C^-_2 s^0_5 s^0_6 d^+_3 d^-_3  +
C^-_2 s^0_7 s^0_8 d^+_3 d^-_3  +
C^-_2 s^+_2 d^+_3 s^-_2 d^-_3  \nonumber \\
& + C^-_2 d^+_2 s^+_3 s^-_2 d^-_3  +
C^-_2 s^0_5 d^-_2 s^-_2 d^-_3  +
C^-_2 s^0_8 s^-_2 d^-_3 d^-_3  +
C^-_2 s^+_2 d^+_3 d^-_2 s^-_3  \nonumber \\
& + C^-_2 d^+_2 s^+_3 d^-_2 s^-_3  +
C^-_2 s^0_5 d^-_2 d^-_2 s^-_3  +
C^-_2 d^+_3 s^+_3 d^-_3 s^-_3  +
C^-_2 s^0_8 d^-_2 d^-_3 s^-_3  \nonumber \\
\end{align}
\begin{align}
F_3 =\ &
C^0_6 C^-_2 s^0_5 s^0_6 h_2  +
C^0_6 C^-_2 s^0_7 s^0_8 h_2  +
C^0_5 D^-_2 s^+_3 d^-_2 h_2  +
C^0_6 C^-_2 s^+_2 s^-_2 h_2  \nonumber \\
& + C^0_6 C^-_2 s^+_3 s^-_3 h_2  +
C^0_5 C^-_2 s^0_6 s^0_8 h_4  +
C^0_6 D^-_1 s^+_2 d^-_2 h_4  +
C^0_5 C^-_2 s^+_3 s^-_2 h_4  \nonumber \\
& + C^0_6 D^-_1 s^+_3 d^-_3 h_4  +
C^0_6 C^-_2 s^0_5 s^0_7 \bar{h}_2 +
C^0_5 D^-_2 s^+_2 d^-_2 \bar{h}_2 +
C^0_5 D^-_2 s^+_3 d^-_3 \bar{h}_2 \nonumber \\
& + C^0_6 C^-_2 s^+_2 s^-_3 \bar{h}_2 +
C^0_5 C^-_2 s^0_5 s^0_6 \bar{h}_4 +
C^0_5 C^-_2 s^0_7 s^0_8 \bar{h}_4 +
C^0_5 C^-_2 s^+_2 s^-_2 \bar{h}_4 \nonumber \\
& + C^0_6 D^-_1 s^+_2 d^-_3 \bar{h}_4 +
C^0_5 C^-_2 s^+_3 s^-_3 \bar{h}_4 +
C^0_5 d^+_3 d^-_2 h_1   C^-_3
+ C^0_6 d^+_2 d^-_2 h_3   C^-_3  \nonumber \\
&+C^0_6 d^+_3 d^-_3 h_3   C^-_3 +
C^0_6 C^0_6 h_2 h_3   C^-_3 +
C^0_5 d^+_2 d^-_2 \bar{h}_1   C^-_3 +
C^0_5 d^+_3 d^-_3 \bar{h}_1   C^-_3 \nonumber \\
& + C^0_5 C^0_6 h_2 \bar{h}_1   C^-_3
 + C^0_5 C^0_6 h_1 \bar{h}_2   C^-_3 +
C^0_6 d^+_2 d^-_3 \bar{h}_3   C^-_3 +
C^0_5 C^0_6 h_4 \bar{h}_3   C^-_3  \nonumber \\
&+ C^0_5 C^0_6 h_3 \bar{h}_4   C^-_3
+ C^0_5 C^0_5 \bar{h}_1 \bar{h}_4   C^-_3.
\end{align}

\subsection{SM Yukawa couplings}\label{SMYukawa}

\subsubsection{$Q_{\rm em}=\frac23$ quark masses}
In the $\overline{\bf 10}_i {\bf 5}_j {\bf 5}_H$ couplings of Table
\ref{yukawas}, we can read off that the following singlet field
must develop near GUT scale VEVs to have a realistic $Q_{\rm em}=\frac{2}{3}$
mass matrix,
\begin{equation}
h_1, \bar{h}_1, h_3, \bar{h}_3, C_5^0, C_6^0.
\end{equation}
To achieve the hierarchy $\frac{m_u}{m_t} \sim 10^{-6}$ among up-type quarks,
some of the singlet vev might also have hierarchically small values $\sim
10^{-3} M_{\rm st}$.

\subsubsection{$Q_{\rm em}=-\frac13$ quark masses}
 According to Table \ref{yukawas}, the
following singlets are required to develop near GUT scale VEVs
toward a successful $Q_{\rm em}=-\frac13$ mass matrix,
\begin{equation}
h_1,\bar h_1, h_3, \bar h_3.
\end{equation}
These fields can have GUT scale VEVs and by some fine tuning we can
obtain realistic $Q_{\rm em}=-\frac13$ quark mass matrix.

\subsubsection{$Q_{\rm em}=-1$ lepton masses}
\label{leptonmasses}
Finally consider the charged lepton mass matrix. According ${\bf 5}_i {\bf 1}_j
\bar {\bf 5}_H$ couplings of Table \ref{yukawas}, the following fields are
likely to have some VEVs of order $M_{\rm GUT}\sim M_I$:
\begin{eqnarray}
\begin{array}{c}
 h_1,\bar{h}_1, h_3, \bar{h}_3,\\
s^0_2
 \end{array} \label{highscalevevfield}
\end{eqnarray}
Out of these, $s^0_2$ is required not to obtain a GUT scale VEV but
a somewhat smaller scale at $M_{LI}$. The GUT scale VEVs themselves
may have a small hierarchy to fit to the experimental values for
masses and mixing angles. The reason for choosing a smaller VEV for
$s_2^0$ is the following. The charged lepton matrix ${\bf 5}_i {\bf 1}_j
\bar {\bf 5}_H$
from Table \ref{Superpotentials} is proportional to $\langle
5_H\rangle$ times
\begin{eqnarray}
 \tilde{M}_e \sim \left(
\begin{array}{ccc}
 0  & 0 & 0  \\
 0 & h_3 h_3 & 1 + \bar{h}_1 h_3 + h_1 \bar{h}_3 \\
0 & 1+ \bar{h}_1 h_3 + h_1 \bar{h}_3 & \bar{h}_1 \bar{h}_1
\end{array}
\right) + \left(
\begin{array}{ccc}
s^0_2 s^0_2 & s^0_2 h_3 & s^0_2 \bar{h}_1 \\
s^0_2 h_3 & 0 & 0 \\
s^0_2 \bar{h}_1 & 0 & 0
\end{array}
\right),
\end{eqnarray}
where we separate the mass matrix into the leading term and the
perturbation in powers of $s^0_2$. Eigenvectors of the left $s^0_2$
independent matrix are $(1,0,0)^T$ with eigenvalue zero\footnote{The
state  $(1,0,0)^T$ will be interpreted as the electron.} and the
others which has the component only in the second and the third
elements.  This implies that the change of eigenvalues for the
second and the third eigenstates are vanishing in the first order
perturbation. Three eigenvalues in this order are given by
\begin{eqnarray}
m_e &=& (s^0_2)^2 \nonumber \\
m_\mu &=&\textstyle \frac{1}{2} \left( h_3 h_3 + \bar{h}_1 \bar{h}_1
- \sqrt{(h_3 h_3 - \bar{h}_1 \bar{h}_1)^2 +
4( 1+\bar{h}_1 h_3 + h_1 \bar{h}_3 )^2} \right) \\
m_\tau  &=&\textstyle \frac{1}{2} \left( h_3 h_3 + \bar{h}_1
\bar{h}_1 + \sqrt{(h_3 h_3 - \bar{h}_1 \bar{h}_1)^2 + 4( 1+\bar{h}_1
h_3 + h_1 \bar{h}_3 )^2} \right)\nonumber
\end{eqnarray}
However, for $m_e$, the second order perturbation will also give
$(s^0_2)^2$ order contribution, so we have to consider
$O((s^0_2)^2)$ for an accurate estimate of electron mass.

\subsubsection{$\mu$ term}\label{subsub:mu}
Finally, let us examine the $\mu$-term among the relevant low energy
couplings. Since the $\mu$ parameter must be of order the
electroweak scale which is negligible compared to the string scale,
the $\mu$ parameter is presumably protected by a global symmetry
\cite{KimNilles93}. Under \Uan\ and \UglA, $T4_5 U2_{\bar{5}}$ is
not invariant, and hence the $\mu$-term is generated by some
symmetry breaking VEVs of  \Uan\ and \UglA. Up to dimension 7, we
find that the $\mu$ term related couplings are  the following
\begin{eqnarray}
 T4_5 U2_{\bar{5}} \times \left\{
\begin{array}{c}
  s^0_2 s^0_4, s^u s^0_2 s^0_2, s^0_2 s^0_4 (h_3 \bar{h}_1 + h_1 \bar{h}_3 ),
  s^u s^0_2 s^0_2 ( h_3 \bar{h}_1 + h_1 \bar{h}_3 ), \\
  D^+_1 D^-_1 s^0_4 ( h_1 \bar{h}_4 + h_4 \bar{h}_1 ),
  D^+_1 D^-_2 s^0_4 ( h_2 \bar{h}_3 + h_3 \bar{h}_2 ), \\
  C^0_3 s^0_3 (s^0_4 s^0_8 h_1 + d^+_3 d^-_1 h_1 + s^0_4 s^0_5 \bar{h}_1
+ d^+_2 d^-_1 \bar{h}_1 ), \\
  C^0_6 s^0_2 s^0_2 s^0_4 h_3 ,  C^0_3 s^0_1 s^0_4 s^0_8 h_3 ,
 C^0_3 s^0_1 d^+_3 d^-_1 h_3,  \\
  C^0_5 s^0_2 s^0_2 s^0_4 \bar{h}_1, C^0_3 s^0_1 s^0_4 s^0_5 \bar{h}_3,
 C^0_3 s^0_1 d^+_2 d^-_1 \bar{h}_3
\end{array}
\right\}
\end{eqnarray}
These $\mu$ terms satisfy the \Uan\ symmetry. Since $T4_5
U2_{\bar{5}} ({\rm i.e.}H_uH_d)$ carries 8 units of the \Uan\
charge, the multiplied factor carries --8 units of \Uan\ charge. If
the singlet fields multiplied are flat directions, we can determine
their VEVs by the minimization. For simplicity, we discuss the first
two terms. Since the VEV of $T4_5 U2_{\bar{5}}$ is nonzero, the
factor $s_2^0 s_4^0+s^u(s_2^0)^2$ must choose the minimum. If the
VEVs of $s_2^0, s_4^0$ and $s^u$ is not determined, probably being
flat directions, this $\mu$ term determines a relation between
$s_2^0, s_4^0$ and $s^u$ through the minimization condition. It is
$s_2^0=-s_4^0/2s^u$ with the result $\mu=-(s_4^0)^2/4s^u$.
  Since the electron mass is
$\sim (s_2^0)^2$, we take  $s^0_2 \sim 10^{15}$ GeV. Then, the
electroweak scale $\mu$ is obtained for example for $s^0_4 \sim
10^{5}$ GeV with an intermediate scale $s^u \sim 10^{12}$ GeV. Of
course, if they are not flat directions, their VEVs are determined
by other more important terms.

\subsection{Search for an approximate symmetry}
 For the sake of showing the existencee of vacua, let us choose
only two kinds of VEVs, one kind carrying only the \Uan\ and the
other kind carrying only the \UglA\ charge. In this way, we can
separate two axion scales.

From the Yukawa couplings and mass terms for the vectorlike pairs,
we require near-GUT-scale VEVs to the fields:
\begin{align}
h_1(0,0),\, \bar h_1(0,0) ,\, h_3(0,0), \, \bar h_3(0,0),\nonumber \\
 C^0_3(-3,0),\, C_5^0(4,0),\, C_6^0(4,0),
\nonumber \\
s^0_2(-4,0),\, s^-_1(-6,0), \nonumber \\
\left( s^+_2(5,0)\mbox{~or~}s^+_3(5,0) \right), \nonumber \\
\left(s^0_1(-4,0)\mbox{~or~}s^0_3(-4,0) \right),  \label{largeVEVs}
\end{align}
where  $(Q_{\rm an},Q_{\rm glA})$ charges are shown.  VEVs of these
fields break \Uan\ only. To break \UglA, we can consider
the following
\begin{equation}
s^u(0,4),s_4^0(-4,4), s_6^0(1,-4).\label{smallVEVs}
\end{equation}
Thus, the domain wall numbers of the U(1)$_{\rm an}$ and U(1)$_{\rm
glA}$ axions are 9 and 1, respectively.


\section{Axion-Photon-Photon Coupling}\label{sec:agg}

The PQ mechanism employs a global symmetry. The $R$ symmetry is not
considered for a Peccei-Quinn mechanism. This is because if we try
to embed the axion field in the phase of a scalar field then this
phase does not carry a SUSY parameter. So for the PQ mechanism
gauginos are neutral.

Let us consider the QCD axion and one hidden sector axion
corresponding to SO(10)$^\prime$. When we have two axions, we must
consider the axion mixing and the cross theorem
\cite{Kim:1998kx,Kim:2006aq}. The higher axion potential corresponds to a
lower axion decay constant.

As discussed in the previous section, we have GUT scale VEVs for the
fields of (\ref{largeVEVs}) and intermediate scale VEVs $M_I$ for
the fields of (\ref{smallVEVs}). In fact, $s_2^0$ and some other
fields which appear in up-type Yukawa couplings have the VEV scale
$M_{LI}$ somewhat larger than $M_I$ but smaller than $M_{\rm GUT}$
for a successful electron mass and for a ratio of up and top quark
masses. The intermediate scale VEVs are in the region $\sim 10^{11}$
GeV. So, we study the following case,
\begin{equation}
 F_A\sim 10^{16}{\rm GeV},\quad F_G\sim 10^{11}{\rm GeV}
\end{equation}
where $F_A$ and $F_G$ correspond to axion decay constants derived
from spontaneous symmetry breaking of \Uan\ and \UglA, respectively.
It turns out that the decay constant of QCD axion is made to be of
order GUT scale.

The confining group we are interested in is
SU(3)$_c\times$SO(10)$^\prime$. We assume that a subgroup of
SO(10)$^\prime$ such as SU(4) or SO(8) confines around $10^{11-13}$
GeV.

The early example of the axion mixing is between the MI axion and
the MD breathing mode axion coupling as \cite{Choi:1985bz},
$$
\frac{1}{32\pi^2F_{\rm MI}}a_{A}(F\tilde F+F^\prime\tilde
F^\prime)+\frac{1}{32\pi^2F_{\rm MD}}a_{G}( F\tilde F-
F^\prime\tilde F^\prime).
$$

But in our case the couplings are more complicated. The phases of
$C_3^0(-3,0)$, $C_5^0(4,0)$, $C_6^0(4,0)$, and
$s^u(0,4),s_4^0(-4,4)$, $s_6^0(1,-4)$, for example couple
according to their \Uan\ and \UglA\ charges. The axion decay
constant can be calculated following the discussion of Sec.
\ref{sec:axion}. Most probably, both axions, the QCD axion and the
hidden sector axion, have the decay constants at the GUT scale
because of the constraints on Yukawa couplings.

Here, for the simplicity of illustration with relatively complex
\Uan\ and \UglA\ charges, let us choose VEVs of $\langle
C_6^0(4,0)\rangle\equiv V_1$ and $\langle s_4^0(-4,4)\rangle\equiv
V_2$ for global symmetry breaking. The relation $s_4^0\simeq
-2s^us_2^0$ of Subsubsec. \ref{subsub:mu} has been an illustrative
example for flat directions, and here we present the coupling
calculation to show the validity of cross theorem and the
calculational method of axion-photon-photon coupling with a field
having both global charges. So the VEV scale of $s_4^0$ is chosen at
an arbitrary scale here.  The U(1)$_{\rm an}$ breaking direction
with $C_6^0$ and $s_4^0$ is
\begin{equation}
\sqrt{4^2V_1^2+4^2V_2^2}\equiv \tilde V
\end{equation}
and the U(1)$_{\rm glA}$ breaking direction is
\begin{equation}
\sqrt{4^2V_2^2}=4V_2.
\end{equation}

 The normalized phase component for the \Uan\ symmetry is
\begin{equation}
a_{\rm an}=\frac{4V_1a_C-4V_2a_s}{\sqrt{4^2V_1^2+4^2V_2^2}}
=\cos\theta_{\rm an}a_C-\sin\theta_{\rm an}a_s.\label{axanomdef}
\end{equation}
The component orthogonal to $a_{\rm an}$ can be written as
\begin{equation}
a_{\rm ortho}=\sin\theta_{\rm an}a_C+\cos\theta_{\rm an}a_s,
\end{equation}
or $a_C$ and $a_s$ are given by
\begin{equation}
a_C=\cos\theta_{\rm an}a_{\rm an}+\sin\theta_{\rm an}a_{\rm
ortho},\quad a_s=-\sin\theta_{\rm an}a_{\rm an}+\cos\theta_{\rm
an}a_{\rm ortho}.\label{defortho}
\end{equation}
In the $V_1\gg V_2$ limit, we have
\begin{equation}
\cos\theta_{\rm an}\simeq 1,\quad \sin\theta_{\rm an}\simeq
\frac{V_2}{V_1}
\end{equation}
The phase component for the \UglA\ symmetry is
\begin{equation}
a_{\rm glA}={\rm the\ phase\ of}\ s_4^0=a_s.
\end{equation}

Since U(1)$_{\rm an}$--SU(5)--SU(5) and U(1)$_{\rm
an}$--SO(10)$'$--SO(10)$'$ anomalies are --9 units, and U(1)$_{\rm
glA}$--SU(5)--SU(5) is 0 and U(1)$_{\rm glA}$--SO(10)$'$--SO(10)$'$
anomaly is {4}, we obtain the following couplings
\begin{align}
 &\left({-9}\frac{a_{\rm an}}{\tilde V}
\right)\frac{1}{32\pi^2}(F\tilde F+F^\prime\tilde F^\prime) +
\left({4}\frac{a_{\rm glA}}{4V_2} \right)
 \frac{1}{32\pi^2}(F'\tilde F')\nonumber\\
&=\frac{1}{32\pi^2} \left({-9}\frac{a_{\rm an}}{\tilde
V}\right)F\tilde F +\frac{1}{32\pi^2}\left( {-9}\frac{a_{\rm
an}}{\tilde V}+ {4}\frac{a_{\rm glA}}{4V_2} \right)F^\prime\tilde
F^\prime . \label{axioncouplfield}
\end{align}
from which we obtain a potential proportional to
\begin{align}
-\Lambda_{QCD}^4 &\cos\left( \frac{a_{\rm an}}{\tilde
V/{9}}\right)-\Lambda_{h}^3m\cos\left( \frac{a_{\rm an}}{\tilde
V/9}-\frac{a_{\rm glA}}{{V_2}}
\right)\nonumber\\
&\simeq-\Lambda_{QCD}^4 \cos\left(\frac{1}{F_1}\tilde a_C
\right)-\Lambda_{h}^3m\cos\left( \frac{1}{F_1}\tilde a_C-
\frac{1}{F_2}\tilde a_{s}\right) \label{axioncoupl}
\end{align}
where we used
\begin{equation}
F_1=\frac{V_1}{9},\quad F_2=\frac{V_2}{4},\quad a_C=4\tilde
a_C,\quad a_s=4\tilde a_s
\end{equation}
and took the limit
$$
\tilde V\gg V_2.
$$
Thus, the hidden sector axion decay constant is
\begin{equation}
F_{h}=\frac{V_2}{4}.\label{QCDdecay}
\end{equation}

>From Eq. (\ref{axioncoupl}), the axion mass matrix is estimated as
\begin{equation}
M^2\propto \left( \begin{array}{cc}
\frac{\Lambda_{QCD}^4}{F_1^2}+\frac{m\Lambda_h^3}{F_1^2}, &
-\frac{m\Lambda_h^3}{F_1F_2}
 \\
 -\frac{m\Lambda_h^3}{F_1F_2},
 &\frac{{m\Lambda_{h}^3}}{F_2^2}
\end{array}
 \right)
\end{equation}
where we simplified the confining scales as $\Lambda_{QCD}$ and
$\Lambda_{h}$, and $m$ is the mass scale of the hidden sector
quarks. Introducing hierarchically small parameters
\begin{equation}
\epsilon=\frac{F_2^2}{F_1^2},\quad\eta=
\frac{\Lambda_{QCD}^4}{m\Lambda_h^3},
\end{equation}
we obtain the larger hidden sector axion mass and the smaller QCD
axion mass
\begin{equation}
\begin{array}{l}
m^2_h\simeq\frac{m\Lambda_h^3}{F_2^2}\\
m^2_{a,\rm QCD}\simeq \frac{\Lambda_{QCD}^4}{F_1^2}
\end{array}
\end{equation}
which shows the validity of the cross theorem on the axion masses
and decay constants.


{ Until now in this section, we treat $U(1)_{\rm glA}$ as an exact
global symmetry. As mentioned, this symmetry is broken at order
dimension 8 in the superpotential and thus we have to discuss the
effect of its breaking on the axion. With the symmetry breaking
terms, the vacuum may be shifted
to a CP-violating one. 
The potential becomes
\begin{align}
&-\Lambda_{QCD}^4 \cos\left(\frac{a_{\rm an}}{F_1}\right)
-\Lambda_{h}^3m\cos\left( \frac{a_{\rm an}}{F_1}+
\frac{a_{\rm glA}}{F_2}\right)  + \frac{(\delta m)^2}{2}
\left( a_{\rm glA} - \theta_{\rm br} F_1 \right)^2,
\end{align}
where we denote the magnitude of the symmetry breaking term
by $(\delta m)^2$ and the shift of the vacuum angle by $\theta_{\rm br}$.
In the approximation that the shift is small, we linearize
$\partial V / \partial a = 0$. Then,
\begin{align}
\left(
\begin{array}{cc}
\frac{\Lambda^4}{F^2_1} + \frac{m \Lambda_h^3}{F^2_1}  &
\frac{\Lambda^3 m}{F_1 F_2}  \\
\frac{\Lambda_h^3 m}{F_1 F_2} & \frac{\Lambda_h^3 m }{ F^2_2 } + (\delta m)^2
\end{array}
\right) \left(
\begin{array}{c}
a_{\rm an} \\ a_{\rm glA}
\end{array}
 \right) =  \left(
\begin{array}{c}
0 \\
(\delta m)^2  \theta_{\rm br} F_2
\end{array}
\right).
\end{align}
We have two cases : (i) $\delta m^2 \gg \frac{\Lambda_h^3 m}{F_2^2} $,
(ii) $\delta m^2 \ll \frac{\Lambda_h^3 m}{F_2^2} $.
For the case (i), the QCD vacuum angle shift is
\begin{eqnarray}
\frac{a_{\rm an}}{F_1} \sim -\theta_{\rm br}.
\end{eqnarray}
So this cannot be used as a QCD axion. But if it belongs to the case
(ii), we have
\begin{eqnarray}
\frac{a_{\rm an}}{F_1} \sim -
\frac{(\delta m)^2}{\frac{\Lambda^3 m}{F_2^2}}
\theta_{\rm br}
\end{eqnarray}
which can be sufficiently small. }

Let us now proceed to calculate the axion--photon--photon coupling.
The global anomaly is
\begin{align}
&{\rm U(1)}_{\rm an}-F_{\rm em}^{\mu\nu}-\tilde F^{\rm em}_{\mu\nu}
:\quad -15\\
&{\rm U(1)}_{\rm glA}-F_{\rm em}^{\mu\nu}-\tilde F^{\rm em}_{\mu\nu}
:\textstyle\quad\quad 0
\end{align}
We choose \Uan\ and \UglA\ for the axions. For illustration we
present again for the case of $C_6^0$ and $s_4^0$.

\FIGURE{
\centering \epsfig{figure=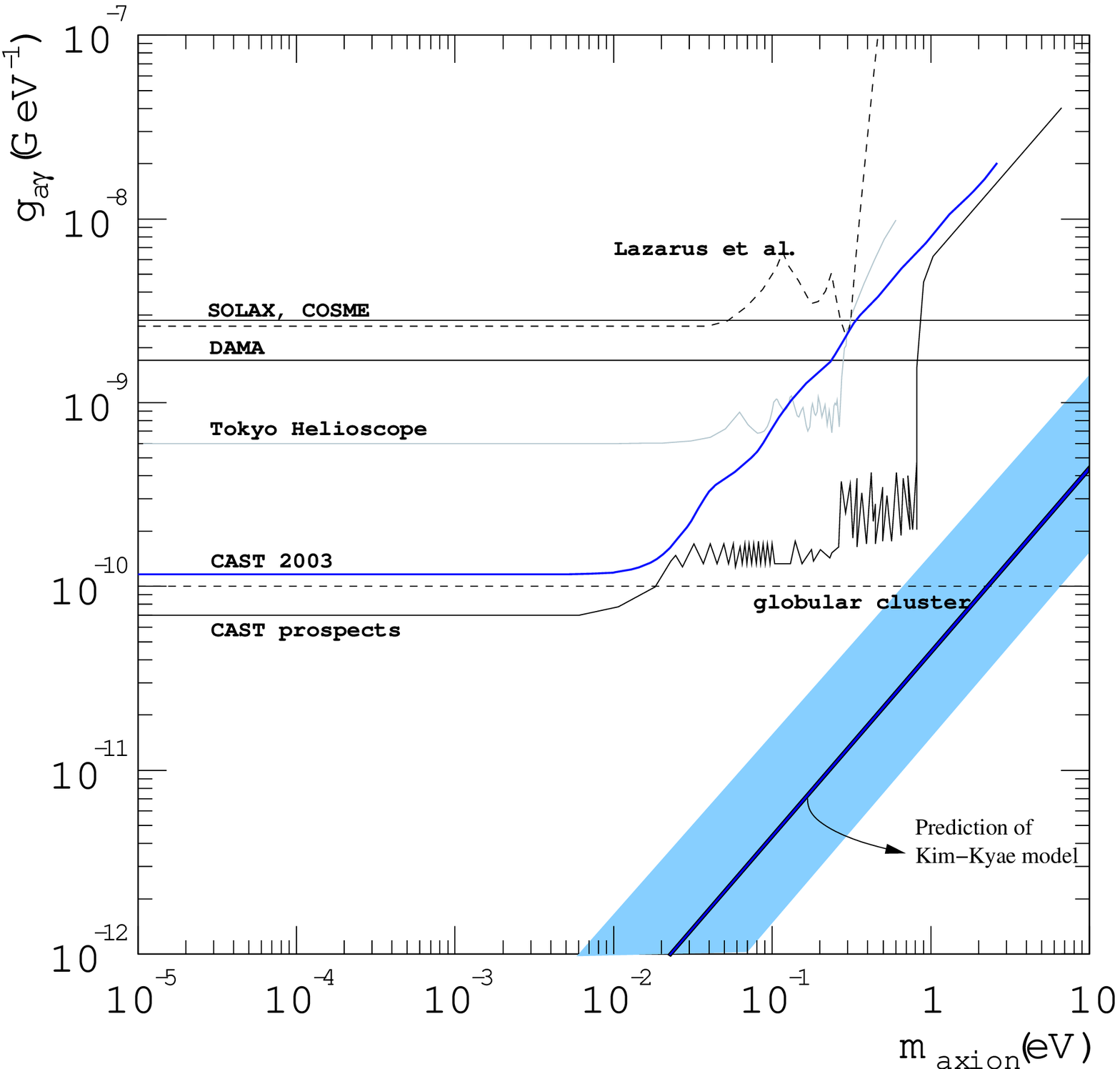, width=15cm}
\caption{\label{axionexp} Experimental bound from various
experiments, especially including CAST 2003 data. This figure is
taken from the result paper of CAST experiment
\cite{Zioutas:2004hi}. Here,we show the prediction of ${\bf Z}_{12}$ model
by the thick line. The blue band is the 20 \% theoretical error of
Ref. \cite{Kaplan:1986ru}. }
}

 For the electromagnetic field $F_{\rm
em}^{\mu\nu}$, we have the following anomaly
\begin{align}
\frac{1}{32\pi^2}&
 \left(\frac{{-15}a_{\rm an}}{\tilde V}
 F_{\rm em}\tilde F_{\rm em}
\right)\nonumber\\
&  \simeq\frac{1}{32\pi^2}
 \left(\frac{{-15}(4\tilde a_C+\frac{16}{9}
 \sqrt{\epsilon}\tilde a_s
 )}{4V_1}
 \right)F_{\rm em}\tilde F_{\rm
em}\simeq \frac{1}{32\pi^2}
 \left(\frac{{-15}\tilde a_C}{V_1}
 \right)F_{\rm em}\tilde F_{\rm
em}\\
&=-\frac{1}{32\pi^2}\left(\frac{5}{3}\frac{a_{\rm
QCD}}{F_1}\right)F_{\rm em}\tilde F_{\rm em}\label{aggco}
\end{align}
whence we obtain the following high energy value for the
axion-photon-photon coupling,
 \begin{align}
{\cal L}_{a\gamma\gamma}=\tilde c_{a\gamma\gamma}\frac{a_{\rm
QCD}}{F_{a,\rm QCD}} \left(\frac{e^2}{32\pi^2} F_{\rm em}\tilde
F_{\rm em}\right)
\end{align}
where
\begin{equation}
\tilde c_{a\gamma\gamma}=\textstyle\frac{5}{3}, \quad F_{a,\rm
QCD}=F_1.\label{cggtilde}
\end{equation}
The axion-photon-photon coupling at low energy takes into account
the QCD chiral symmetry breaking  and we obtain
\begin{equation}
c_{a\gamma\gamma}=\tilde c_{a\gamma\gamma}-c_{\chi SB} \simeq -0.26
\end{equation}
where we take $c_{\chi SB}\simeq 1.93$. The value $c_{\chi SB}$
depends on the current quark mass ratio $Z=m_u/m_d$, the instanton
contribution factor $\xi=m_s/\Lambda_{\rm instanton}$, and
$r_{ds}=m_d/m_s$. For $\xi=0.1$ and $r_{ds}=\frac1{20}$, we obtain
\begin{equation}
c_{\chi SB}=\frac{\frac23(4+1.05Z)}{1+1.05 Z}.
\end{equation}
Inclusion of one-loop effects and second order chiral perturbation
give sizable contributions to squared meson masses \cite{Kaplan:1986ru}.
For $Z=\frac59$, we obtain $c_{\chi SB}=1.93$.  With the next order
effect up to the 20\% level, we obtain a band for $c_{\chi SB}$,
i.e. $c_{\chi SB}=1.76\sim 2.26$. Thus, $c_{a\gamma\gamma}$ turns
out to be sizable, which can be detected. Fig. \ref{axionexp} shows
the current experimental search limit on $c_{a\gamma\gamma}$ and the
prediction point of the present model with the band shown. The solid
line corresponds to $c_{\chi SB}=1.93$. It is the first reliable
calculation of the axion-photon-photon coupling. The cavity
detectors \cite{DePanfilis:1987dk} and the high $Z$ Rydberg atom Kyoto axion
detector \cite{KyotoDet} already used this kind of
axion-photon-photon interaction. But the recent CAST detector
\cite{Zioutas:2004hi} seems to be the most promising one for
detection of a very light axion in the region $F_a\sim 10^{11}$ GeV.
The detectability of axion using the $c_{a\gamma\gamma}$ coupling
was proposed by Sikivie \cite{Sikivie:1983ip}. However, the magnitude of
$F_{a,\rm QCD}$ in the model we studied is at the GUT scale and it
cannot be detected in these kinds of detectors.

What will be the case if we use VEVs of all the fields instead of
just $C_6^0$ and $s_4^0$? We have discussed at length that it was
not possible to find a global symmetry which has the QCD anomaly
consistently with the breaking scheme of the flipped SU(5). More
importantly, the hidden sector axion potential is higher than the
QCD axion potential and the cross theorem dictates that the decay
constant of the QCD axion is the breaking scale of \Uan. Thus, in
our model the QCD axion cannot be made detectable.

Finally, we mention that the hidden sector axion potential can be
made smaller than the QCD axion potential by making the hidden
sector quark mass extremely small in which case the cross theorem
acts in the other direction. In this case, we can expect that the
QCD axion can fall in the detectable region. However, we have not
found any approximate global symmetry possessing SU(5) anomaly and
hence this desirable scenario is not realized in the present model.

\section{Conclusion}\label{sec:Conc}

In this paper, we presented a general method to house a QCD axion in
string-derived MSSM models. One related objective is to make it
observable in ongoing or future axion search experiments since the
axion derived from superstring might be the most significant
prediction of string. We presented the criteria that should be
satisfied in these models. Since the magnitude of the axion decay
constant is essential in the solution of the strong CP problem,
cosmology, and in axion search experiments, we presented a general
formula for the decay constant in Eqs. (\ref{DWphaseN}) and
(\ref{AxDecay}). We must consider the full Yukawa coupling structure
of matter fields toward this objective. So far, this kind of full
Yukawa coupling structure has not been studied except in a recent
$\Z_{12-I}$ model \cite{Kim:2006hv}. Here the Yukawa coupling has been
given completely, which made it possible for us to pick up an
approximate global symmetry so that a phenomenologically allowed QCD
axion results. However, in the $\Z_{12-I}$ model we study there does
not exists a vacuum where $F_a\sim 10^{11}$ GeV with successful MSSM
phenomenology. This might be a most probable situation in string
models. It will be very interesting if one can find a string model
with an observable QCD axion.

\acknowledgments{
The authors are very thankful to B. Kyae for invaluable comments and
pointing out some error in the computer program.
This work is supported in part by the KRF ABRL
Grant No. R14-2003-012-01001-0. J.E.K. is also supported in part by
the KRF grants, No. R02-2004-000-10149-0 and  No.
KRF-2005-084-C00001.K.S.C is also supportd in part by the
European Union 6th framework program MRTN-CT-2004-503069
"Quest for unification", MRTN-CT-2004-005104 "ForcesUniverse",
MRTN-CT-2006-035863 "UniverseNet" and
SFB-Transregio 33 "The Dark Universe" by Deutsche
Forschungsgemeinschaft (DFG).
}

\end{document}